\documentclass[12pt,reqno]{amsart}

\usepackage[english]{babel}

\usepackage{cite} 
\usepackage{hyperref}
\hypersetup{colorlinks=true,linkcolor=magenta,anchorcolor=green,citecolor=cyan,filecolor=black,menucolor=black,urlcolor=black}

\usepackage{tikz}
\usetikzlibrary{calc,decorations.pathreplacing}

\usepackage{amsfonts}
\usepackage{amsmath}
\usepackage{amssymb}
\usepackage{latexsym}
\usepackage{cite}

\numberwithin{equation}{section}

\newcount\hh
\newcount\mm
\mm=\time
\hh=\time
\divide\hh by 60
\divide\mm by 60
\multiply\mm by 60
\mm=-\mm
\advance\mm by \time
\def\thistime{\number\hh:\ifnum\mm<10{}0\fi\number\mm}
\newcommand{\bea}{\begin{eqnarray}}
\newcommand{\eea}{\end{eqnarray}}
\newcommand{\be}{\begin{eqnarray}}
\newcommand{\ee}{\end{eqnarray}}

\newcommand{\boundellipse}[3]% center, xdim, ydim
{(#1) ellipse (#2 and #3)
}

\numberwithin{equation}{section}

\def\nn{\nonumber}

\def\Ree{\Re\textrm{e}}
\def\Imm{\Im\textrm{m}}

\def\Li_#1(#2){\textrm{Li}_{#1}\left(#2\right)}
\def\cLi_#1(#2){\mathcal{L}_{#1}\left(#2\right)}
\def\bLi_#1(#2){\mathbf{L}_{#1}\left(#2\right)}

\def\cB{{\cal B}}

\def\cD{{\cal D}}

\def\cJ{{\mathcal J}}
\def\cK{{\mathcal K}}

\def\cM{{\cal M}}

\def\ba{{\bf a}}
\def\bb{{\bf b}}

\def\bx{{\bf x}}
\def\by{{\bf y}}
\def\bz{{\bf z}}

\def\mC{\mathfrak{C}}

\def\mm{\mathfrak{m}}

\def\mp{\mathfrak{p}}

\def\p{\partial}
\def\half{{1\over 2}}

\def\no{\nonumber}
\def\sm{\smallskip}

\def\ZZ{{\mathbb Z}}
\def\IC{{\mathbb C}}
\def\IR{{\mathbb R}}

\def\IN{{\mathbb N}}
\def\IQ{{\mathbb Q}}

\def\cM{\mathcal{M}}
\def\cD{\mathcal{D}}

\def\cB{\mathcal{B}}

\def\pol{polylogarithm}
\def\half{{1 \over 2}}
\def\p{\partial}

\def\Claim{Conjecture}
\def\claim{Conjecture}

\newlength{\abstractwidth}
\title[]{\bf  Modular graph functions}

\author[E. D'Hoker]{Eric  D'Hoker}
\address{Eric  D'Hoker\\ Mani L. Bhaumik Institute for Theoretical Physics,  Department of Physics and Astronomy \\
 University of California, Los Angeles, CA 90095, USA}
\email{dhoker@physics.ucla.edu}

\author[M.B. Green]{Michael B. Green}
\address{Michael B. Green\\ Department of Applied Mathematics and Theoretical Physics\\
Wilberforce Road, Cambridge CB3 0WA, UK}
\email{M.B.Green@damtp.cam.ac.uk}

\author[\"O. G\"urdo\u gan]{\"Omer G\"urdo\u gan}
\address{\"Omer G\"urdo\u gan\\
Institut de Physique Th{\'e}orique\\
CEA, IPhT, F-91191 Gif-sur-Yvette, France\\
CNRS, URA 2306, F-91191 Gif-sur-Yvette, France\hfill\break
Laboratoire de Physique Th\'eorique de l'\'Ecole Normale Sup\'erieure,\\
24, rue Lhomond, 75231 Paris cedex, France}
\email{omer.gurdogan@cea.fr}

\author[P. Vanhove]{Pierre Vanhove}
 \address{Pierre Vanhove\\
 Institut des Hautes Etudes Scientifiques\\
 Le Bois-Marie, 35 route de Chartres\\
 F-91440 Bures-sur-Yvette, France\hfill\break
Institut de Physique Th{\'e}orique\\
CEA, IPhT, F-91191 Gif-sur-Yvette, France\\
CNRS, URA 2306, F-91191 Gif-sur-Yvette, France\hfill\break
Department of Applied Mathematics and Theoretical Physics\\
 Wilberforce Road, Cambridge CB3 0WA, UK}
\email{pierre.vanhove@cea.fr}

\thanks{DAMTP-2015-86, IPhT-t15/202, IHES/P/15/29, LPTENS-15/09}
\date{\today}
\begin{document}

 \begin{abstract}   
In earlier work we studied features of non-holomorphic modular
functions associated with Feynman graphs for a conformal scalar field
theory on a two-dimensional torus with zero external momenta at all
vertices.   Such  functions, which we will refer to as  {\sl modular
  graph functions}, arise, for example,  in the low energy expansion
of genus-one Type II superstring amplitudes.  We here introduce a
class of {\sl single-valued elliptic multiple polylogarithms}, which are defined as elliptic functions associated with Feynman graphs with vanishing external momenta at all but two vertices.  These functions depend on a coordinate, $\zeta$, on the elliptic curve and reduce to modular graph functions when $\zeta$ is set equal to $1$. 
We demonstrate that these single-valued elliptic multiple polylogarithms are linear combinations of multiple polylogarithms, and that modular graph functions are sums of single-valued elliptic multiple polylogarithms evaluated at the identity of the elliptic curve, in both cases with rational coefficients.  This insight suggests the many interrelations between modular graph functions (a few of which were established in earlier papers) may be obtained as a consequence of  identities involving multiple polylogarithms, and explains an earlier observation that the coefficients of the Laurent polynomial at the cusp  are given by rational numbers times single-valued multiple zeta values.
 
%{\bf Draft version \today\ at \thistime}
\end{abstract}
\maketitle
\tableofcontents
%%%%%%%%%%%%%%%%%%%%%%%%%%%%%%%%%%%%%%%%%%%%%%%%%%%%%%%%%%%%%%%%%%
\section{Introduction}

Superstring perturbation theory may be merely an approximation to a complete non-perturbative formulation of string theory, but it already exhibits a remarkably rich mathematical structure.  The perturbative series is given by a topological expansion for two-dimensional surfaces which represent string world-sheets. For the closed superstring theories the perturbative series is  given by the sum over all genera~$g$, with $g\ge 0$,  of functional integrals over orientable  (super)Riemann surfaces.   In the case of four-graviton amplitudes, which will be the prototype and point of departure of this paper, the explicit expressions  for tree-level ($g=0$) and genus-one ($g=1$) were obtained in \cite{Green:1981yb}, for genus-two ($g=2$)  in \cite{D'Hoker:2005jc},  and a certain amount is known about the leading low energy behaviour of the genus-three ($g=3$) case in \cite{Gomez:2013sla}. For the open string theory, a summation over boundaries and cross-caps must also be included.

\smallskip

The low energy expansion of string theory corresponds to an expansion valid when the energies and momenta are small in units of the inverse of the string length scale~$\ell_s$, a parameter which is related to Newton's gravitational constant. 
The lowest order contribution corresponds to Einstein's theory, while
higher order corrections become important for strong gravitational
fields. The structure of these higher order corrections is of
considerable mathematical interest. In particular, their coefficients
in the low energy expansion of tree-level $N$-particle amplitudes  in
open superstring theory are multiple zeta values, which are special
values of multiple polylogarithms.  The analogous coefficients of
tree-level  $N$-particle closed superstring  amplitudes  are
single-valued multiple zetas~\cite{Stieberger:2013wea,Stieberger:2014hba}, which in turn  are special values of single-valued multiple polylogarithms, following the terminology introduced in~\cite{brownCRAS,Schnetz:2013hqa,Brown:2013gia}.  

\smallskip

Much less is understood about the low energy expansion of string
amplitudes with higher genus.  In the genus-one closed superstring
case the coefficients in the low energy expansion are given by
integrals of non-holomorphic modular functions over the complex structure modulus $\tau$ of the torus
that is defined by the string world-sheet.  These modular functions,
which can be expressed as multiple sums, are generalisations of
non-holomorphic Eisenstein series. Specific classes of such
functions have been shown to satisfy a number of very intriguing
relationships \cite{D'Hoker:2015foa,D'Hoker:2015zfa,Basu:2015ayg}.  
These  are reminiscent of the algebraic relationships between multiple zeta values
but, in the present case they are relationships between functions
defined on an elliptic curve. In the genus-two closed superstring case, 
a connection has been uncovered with the Zhang-Kawazumi invariant \cite{D'Hoker:2013eea}, which
satisfies equally intriguing relations \cite{D'Hoker:2014gfa}, but whose study remains incomplete.

\smallskip

In the present paper, we specialise to the case of a conformal scalar field theory on a two-dimensional torus, or elliptic curve, with arbitrary complex modulus $\tau$.
To every Feynman graph with vanishing external momenta on each vertex, $L$ loops, and $w$ scalar Green functions on the edges of the graph, we associate a non-holomorphic modular function in $\tau$ of depth $L$ and weight $w$, which we shall refer to as a {\it modular graph function}. Certain classes of graphs will produce vanishing modular functions, such as any one-edge reducible graph, and any graph in which at least one vertex supports only a single Green function edge. In this paper, we shall consider graphs with non-derivative couplings only, but this restriction can be easily lifted if needed. 

\smallskip

We shall also introduce elliptic functions that depend on a point $\zeta$ on the elliptic curve of modulus $\tau$, and that are associated with Feynman graphs in which all  but two vertices have vanishing external momenta.  These elliptic functions provide examples of {\it single-valued elliptic multiple  polylogarithms}. We will  demonstrate that  any modular graph function may be expressed as the value of a single-valued
elliptic multiple polylogarithm\footnote{The elliptic  multiple polylogarithms considered in this paper are not holomorphic  and differ from those introduced in~\cite{brownlevin}, which are of  relevance to the open string annulus amplitude, as discussed  in~\cite{Broedel:2014vla,Broedel:2015hia}. Although there is a clear
  relationship between the open string and the closed string, this
  relationship is not a subject studied in this paper.} when $\zeta$ 
is set equal to the particular value $\zeta=1$ corresponding to the
identity on the elliptic curve.  This is an elliptic analogue of the
familiar statement that the single-valued multiple-zeta values
discussed in~\cite{brownCRAS,Schnetz:2013hqa,Brown:2013gia} are the
values of single-valued multiple polylogarithms with their arguments
set equal to 1. It is therefore natural to call these special
  values \emph{single-valued elliptic multiple zetas}. This connection between modular graph functions and single-valued elliptic multiple polylogarithms suggests a compelling origin of the many interrelations between modular graph functions (a few of which were motivated in \cite{D'Hoker:2015foa}) as a consequence of  identities involving elliptic polylogarithms.

\subsection{Outline of paper}

Section~\ref{sec:overview}  will give a brief overview of some of the relevant features of polylogarithms, multiple polylogarithms, multiple zeta values, their single-valued projections,  and their elliptic generalisations that will enter into the subsequent ideas in the paper.  Section~\ref{sec:string} will discuss the modular graph functions and single-valued elliptic multiple polylogarithms that arise in the low energy expansion of the perturbative amplitudes in superstring theory and  which are expressed in terms of Feynman graphs for a conformal scalar field theory on a two-dimensional torus. In particular, we will show that every modular graph function is given by a single-valued elliptic multiple polylogarithm evaluated at a special point. Section~\ref{sec:graphs} illustrates  this feature by considering some of the  infinite classes of graphs  studied
in~\cite{Green:1999pv,Green:2008uj,D'Hoker:2015foa,D'Hoker:2015zfa,Basu:2015ayg} by other methods. Section~\ref{sec:svEgraphs}  develops the conjecture which states that single-valued multiple polylogarithms are linear combinations of elliptic polylogarithms with rational coefficients, offers a proof for the infinite class of star graphs, and outlines some of the arguments for general graphs. In section \ref{sec:6}, the validity of the conjecture is shown to lead to a corollary stating  that the non-leading coefficients of the Laurent expansion of the constant Fourier mode of modular graph functions are single-valued multiple zeta values. A summary and further thoughts on the basis of modular graph functions, and their further generalisations,  is relayed to section \ref{sec:highern}.

\sm

Appendix~\ref{sec:covar-deriv} presents some relations between  Eichler integrals, elliptic polylogarithms, and holomorhic Eisenstein series in preparation of appendix~\ref{sec:C111} where  the algebraic properties of multiple polylogarithms will be used to evaluate the modular graph  function associated with the simplest two-loop graph, $C_{1,1,1}(q)$.  This calculation makes use of various reduction identities for multiple sums that are determined in  appendix~\ref{sec:polyred}.

%%%%%%%%%%%%%%%%%%%%%%%%%%%%%%%%%%%%%%%%%%%%%%%%%%%%%%
%%%%%%%%%%%%%%%%%%%%%%%%%%%%%%%%%%%%%%%%%%%%%%%%%%%%%%
\section{Some basic features of multiple polylogarithms}
\label{sec:overview}
%%%%%%%%%%%%%%%%%%%%%%%%%%%%%%%%%%%%%%%%%%%%%%%%%%%%%%
%%%%%%%%%%%%%%%%%%%%%%%%%%%%%%%%%%%%%%%%%%%%%%%%%%%%%%

The emphasis in this paper is the analogy between elliptic functions
that arise as coefficients  in the low energy expansion in genus-one
closed superstring amplitudes  with  single-valued multi-zetas that arise  in the expansion of the tree-level
amplitudes~\cite{Stieberger:2013wea,Stieberger:2014hba}.  For clarity
we will here give a brief (and incomplete) review of the relationship
of these quantities to multiple polylogarithms and to single-valued
multiple polylogarithms.   There are many detailed reviews of this large subject in the literature and we note in particular the elementary introduction in~\cite{Waldschmidt}.

\subsection{Polylogarithms}
\label{sec:mult-polyl-mult}

The  polylogarithm $\Li_a(z)$ is defined for any value of $a \in \IC$ by the power series expansion
\bea
\Li_a(z) = \sum_{k=1}^\infty \frac{z^k}{k^a}\,,
\label{e:polylogdef}
\eea
which is absolutely convergent for $|z|< 1$. The \pol\, is a natural generalisation of the logarithm since we have $\Li_1(z)= - \log(1-z)$. Alternatively, the function $\Li_a(z)$ may be defined by the integral representation
\bea
\Li_a(z) = { z \over \Gamma (a) } \int _1 ^\infty { dt \over t} \, { (\log t)^{a-1} \over t-z} \, ,
\eea
which coincides with (\ref{e:polylogdef}) for $|z|<1$, but may be analytically continued to all $z \in \IC \backslash [1,+\infty[$.
The resulting $\Li_a(z)$ manifestly has a branch point at $z=1$, but it also has a branch point at $z=0$ on its higher Riemann sheets. Therefore, the function $\Li_a(z)$ for generic values $a \in \IC$ is multiple-valued  and has interesting monodromies, again generalising the properties of $\Li_1(z)=-\ln (1-z)$. Of particular interest is the relation of $\Li_a(z)$ to the Riemann zeta function $\zeta (a)$  via
\bea
\Li_a(1) = \zeta(a)\,,
\label{e:specialpoly}
\eea
and to the Bernoulli polynomials $B_n(x)$ via
\bea
\Li_n(e^{2i\pi x}) + (-1)^n \Li_n(e^{-2i\pi x}) = - \frac{(2 i\pi)^n}{n!}\, B_n(x)\,,
\label{e:polybern}
\eea
for $n \in \IN$ and $x\in [0,1]$. Recall that $B_n(x)$ is the $n$-th Bernoulli polynomial  which is defined by the expansion in powers of $t$ of the following generating function
\bea
\frac{t \, e^{xt}}{e^t-1}=\sum_{n=0}^\infty B_n(x)\frac{t^n}{n!} \, .
\label{e:bernoullis}
\eea
Moreover, the function $\Li_a(z)$ satisfies the differential relation 
\bea
z \, \frac{\partial \, \Li_a(z)}{\partial z} = \Li_{a-1}(z)\,,
\label{e:diffpoly}
\eea
and the converse integral relation   
\bea
\Li_{a+1}(z) = \int_0^z {dy \over y}  \, \Li_a(y).
\label{e:intrel}
\eea

\subsection{Single-valued polylogarithms}
\label{sec:singlevalued--polyl-mult}

At the cost of giving up holomorphicity in $z$, it is possible to
construct a single-valued \pol \,  associated with  $\Li_a(z)$. The
Bloch--Wigner dilogarithm is the prototype for such single-valued
polylogarithms, and is defined as follows
\bea
D(z) = \Imm\Big(\Li_2(z) + \log(1-z)\, \log|z|\Big)\,,\qquad\quad z\in \IC\backslash \, \{0,1\}\,.
\label{e:dilogdef}
\eea
In this paper we will make use of the generalisations of the
Bloch--Wigner dilogarithm introduced by Zagier
in~\cite{ZagierDm}. They are single-valued (real analytic)
polylogarithms $D_{a,b}(z)$ for  $z\in\IC\backslash [1,\infty[$ and
$a,b \in \IN$, and can be expressed in terms of sums of ordinary
polylogarithms $\Li_k(z)$ for an integer index $k$ by
\begin{multline}
\label{e:Dabx}
  D_{a,b}(z)= (-1)^{a-1}\sum_{k=a}^{a+b-1}  \left(k-1\atop
    a-1\right)\, {(-2\log|z|)^{a+b-1-k}\over (a+b-1-k)!}\, \Li_k(z)\cr
+  (-1)^{b-1}\sum_{k=b}^{a+b-1} \left(k-1\atop
    b-1\right)\, {(-2\log|z|)^{a+b-1-k}\over (a+b-1-k)!}\, (\Li_k(z))^*\,.
\end{multline}
These functions satisfy the complex conjugation relation $D_{a,b}(z)^* = D_{b,a}(z)$, and we define their {\it weight} to be $a+b-1$. In particular, the function $D_{a,a}(z)$ with $a\in \IN$ has weight $2a-1$ and is real and  single-valued on $\IC\backslash\{0,1\}$.  It is given by a finite linear combination of the $\Li_k(z)$-functions   
\begin{equation}
 D_{a,a}(z)= 2\Ree\left((-1)^{a-1}\sum_{k=0}^{a-1} \,
  {k+a-1\choose a-1}\, {(-2\log|z|)^{a-1-k}\over (a-1-k)!}  \, \Li_{a+k}(z)\right)\,,
  \label{e:Daax}
\end{equation}
and will play an important role in the sequel of this paper.
The following are examples at low weights. At weight one, we have,
\bea
D_{1,1}(z) = -2 \log |1-z|^2 \, ,
\eea
which is manifestly single-valued. At weight 2 we have
\begin{eqnarray}
  D_{1,2}(z)&=&2i D(z)+  2 \log|z| \, \log|1-z| \, ,
  \nn\\[0.5ex]
D_{2,1}(z)&=&-2iD(z)+   2 \log|z| \, \log|1-z| \,,
\end{eqnarray}
where $D(z)$ is the Bloch-Wigner dilogarithm of (\ref{e:dilogdef}) which is single-valued on $\IC\backslash\{0,1\}$. At weight 3 we have
\begin{eqnarray}
 D_{1,3}(z)&=& 2 (\log|z|)^2 \, \Li_1(z) -2 \log|z| \, \Li_2(z)+2 \Ree(\Li_3(z)) \, ,
 \nn\\[0.5ex]
D_{2,2}(z)&=& 4 \log|z| \, \Ree(\Li_2(z))-4 \Ree(\Li_3(z))\,.
\end{eqnarray}
It is not difficult to check that these functions are algebraically independent.

\subsection{Multiple polylogarithms}

The multi-variable polylogarithm $\Li_{a_1,\dots,a_r}(z_1,\dots,z_r)$ is referred to as a multiple \pol\, 
and was defined by~\cite{Goncharov:Mpl,BorweinBradleyBroadhurstLisonek}
\begin{equation}\label{e:Mpl}
\Li_{a_1,\dots,a_r}(z_1,\dots,z_r)=\sum_{0<m_1<\dots <m_r}
\prod_{i=1}^r  {z_i^{m_i}\over  m_i^{a_i}}\,,
\end{equation}
with $a_i\in \IN$ for $1\leq i\leq r$, and with $|z_i|\leq 1$ ($1\leq i\leq r-1$) and $|z_r|<1$. When $a_i\in \IN\geq2$ for
all $1\leq i\leq r$ this function is defined for $|z_i|\leq 1$
for $1\leq i\leq r$.
\smallskip

The {\it weight} of a multiple polylogarithm is the sum of its indices
$\sum_{i=1}^r a_i$ and its {\it depth} is the number of indices, $r$.
The classical polylogarithm functions $\Li_a(z)$ are special cases with
depth $r=1$.

\smallskip

The multiple polylogarithms in~\eqref{e:Mpl} satisfy two kinds of differential
 relations. One of these reduces the weight by one but leaves the depth unchanged
 \begin{equation}\label{e:DiffMpl}
  z_i {\partial\over \partial z_i}
\Li_{a_1,\dots,a_r}(z_1,\dots,z_r)=\Li_{a_1,\dots,a_i-1,\dots,
a_r}(z_1,\dots,z_r)\qquad 1\leq i\leq r\,,
\end{equation}
while the other reduces both the weight and the depth by one 
\begin{equation}
  (1-z_r) {\partial\over \partial z_r}
\Li_{a_1,\dots,a_{r-1},1}(z_1,\dots,z_r)=\Li_{a_1,\dots,
a_{r-1}}(z_1,\dots,z_{r-1}z_r)\,.
\end{equation}
Together with the initial conditions
$\Li_{a_1,\dots,a_r}(0,\dots,0)=0$ these differential equations determine the multiple 
polylogarithms by multiple integration. This leads to Chen iterated integrals~\cite{Chen:II} 
which endow the space of multiple polylogarithms with a {\it shuffle algebra}. Since we
will not make use of this construction we refer to~\cite{Waldschmidt} for details and references. 
For additional reference, we note that these properties have been implemented  in the algebraic  
program {\tt Hyperint} by Erik Panzer~\cite{Panzer:2014caa} and in {\tt MPL} by Christian Bogner in~\cite{Bogner:2015nda}.

\smallskip
 
From the series representation one derives the {\it stuffle relations} as
described in~\cite{Moch:2001zr}, and implemented in~\cite{Weinzierl:2002hv,Moch:2005uc}.
The stuffle relation of two multiple polylogarithms is given by
\begin{equation}
\label{stuffle}
  \Li_{a_1,\dots,a_r}(x_1,\dots,x_r)\Li_{b_1,\dots,b_s}(y_1,\dots,y_s)=
  \sum_{k=\textrm{max}(r,s)}^{r+s} \Li_{c_1,\dots,c_k}( z_1,\dots,z_k)\,,
\end{equation}
where the sum is over all the sequences $c(k):=(c_1,\dots,c_k)$ that arise
in the stuffle product of the sequences $a(r):=(a_1,\dots,a_r)$ and
$b(s):=(b_1,\dots,b_s)$.
The stuffle product, denoted $*$, is a commutative product defined recursively by
\begin{itemize}
\item $(a_1,\dots,a_r) \, *\, \{\}= (a_1,\dots,a_r)$  where $\{\}$ is
  the empty sequence
\item $(a_1,\dots,a_r) \,*\, (b_1,\dots,b_s)= a_1\cdot(
  (a_2,\dots,a_r)\, *\, (b_1,\dots,b_s))+ b_1\cdot(
  (a_1,\dots,a_r)\,*\,(b_2,\dots,b_s))+ (a_1+b_1)\cdot(
  (a_2,\dots,a_r)\,*\,(b_2,\dots,b_s))$
\item  $x\cdot (a_1,\dots,a_r)= (x,a_1,\dots,a_r)$ is the
  concatenation operation
\end{itemize}
The argument $z_i$ associated with the index
$c(k)_i=a(k)_i+b(k)_i$ is obtained as follows: 
\begin{itemize}
\item if $a(k)_i\neq0$ and $b(k)_i=0$ then  $z_i=x_i$;
\item if $a(k)_i=0$ and $b(k)_i\neq 0$ then  $z_i=y_i$;
\item if $a(k)_i\neq0$ and $b(k)_i\neq0$ then  $z_i=x_iy_i$.
\end{itemize}
Note that since we have assumed that $a_i \geq 1$ for all $i=1, \dots, r$ and $b_j\geq 1$ for all $j=1,\dots, s$, the case where both $a(k)_i=0$ and $b(k)_i=0$ cannot arise.
\smallskip

We illustrate this with two examples that will be useful later.  The  first example is the product of two depth-1 polylogarithms
\begin{equation}
  \Li_a(x) \Li_b(y)= \Li_{a+b}(xy)+ \Li_{a,b}(x,y)+\Li_{b,a}(y,x)  \,.
\end{equation}
The second example is the product of a depth-2 multiple polylogarithm by a depth-1 polylogarithm
\begin{multline}
  \Li_{a,b}(x,y)  \Li_c(z)=\Li_{a+c,b}(xz,y)+\Li_{a,b+c}(x,yz)
  \cr
+\Li_{a,b,c}(x,y,z) +\Li_{a,c,b}(x,z,y) +\Li_{c,a,b}(z,x,y)  \,. 
\end{multline}
\subsection{Single-valued multiple polylogarithms}

The special case of single-variable multiple polylogarithms, which have the form
\begin{equation}
  \Li_{a_1,\dots,a_r}(z)= \Li_{a_1,\dots,a_r}(1,\dots,1,z)\,,
\end{equation}
 have   monodromies around $z=0$ and $z=1$.
However single valued versions  on $\mathbb C\backslash\{0,1\}$ of  such functions were constructed by
Francis Brown in~\cite{brownCRAS,Brown:2013gia}. They are obtained  by
appropriate linear combinations of product single-variable multiple
polylogarithms and their complex conjugate to define functions on  $\mathbb
C\backslash\{0,1\}$  without monodromies.  These combinations are
examples of {\it single-valued multiple polylogarithms}. 
A construction of single-valued multiple polylogarithms with more than
one variable has been given in~\cite{DelDuca:2016lad}, which 
appeared some months after the first version of this paper
appeared on the arXiv.

\subsection{Multiple zeta values and single-valued multiple zeta values}
\label{sec:mult-polyl-mult}

Multiple zeta values~\cite{ZagierMZV} provide a natural generalisation of Riemann zeta values. 
The multiple zeta function may be defined by the following multiple sums for $\Ree(a_i) > 1$,  
\bea
\zeta(a_1,\dots,a_r)=
\sum_{0<n_1<\dots<n_r} \prod_{i=1}^r {1\over n_i^{a_i}}\,,
\eea
and analytically continued to $a_i \in \IC$. For $a_i \in \IN$, and $a_r >1$, the quantities $\zeta (a_1, \dots, a_r)$ give {\it multiple zeta values}.  The multiple polylogarithm evaluated with all its arguments equal to 1 is a multiple zeta value given by
\begin{equation}
  \Li_{a_1,\dots,a_r}(1,\dots,1)= \zeta(a_1,\dots,a_r)\,,
\end{equation}
when $a_i\in \IN$ with $a_r >1.$\footnote{When $a_r=1$ the expression diverges and has to be regularised, but this case will not concern us in the sequel.}

\smallskip

The basis of multiple zeta values has been understood in terms of
periods on the Riemann sphere with marked points~\cite{Brown:2009qja}.  This interpretation also arises naturally in the analysis of the low energy expansion of string theory tree amplitudes~\cite{SCHLotterer:2012ny,Broedel:2013aza,Broedel:2013tta}. 
The stuffle and shuffle relations for multiple polylogarithms lead
to a wealth of algebraic relations between multiple zeta values.  

\smallskip

The single-valued multiple zetas are obtained by setting the arguments 
of single-valued multiple polylogarithms to 1. For example,
\begin{eqnarray}
D_{1,2}(1)&=&D_{2,1}(1)=0,\cr
D_{1,3}(1)  &=&2\zeta(3)\cr
D_{2,2}(1) &=&-4\zeta(3)\,.
\end{eqnarray}
It is not difficult to check that these functions have  the properties  
\bea
\label{e:Dabprops}
D_{a,b}(1)&=&0\qquad\qquad\qquad\qquad\qquad \  a+b\in 2 \, \IN-1
\nn\\
D_{a,b}(1)&\in& \zeta(a+b-1)\times \ZZ\qquad\qquad a+b\in 2 \, \IN\,. 
\eea 
This has led Brown to define {\it single-valued zeta values} $\zeta _{sv}$ that are given by~\cite{Brown:2013gia} 
\begin{eqnarray}
\zeta_{sv}(2n)&=&0 \hskip 1in  n \in \IN \cr
  \zeta_{sv}(2n+1)&=&2\zeta(2n+1)\,.
\end{eqnarray}

\smallskip

It is important to appreciate that the definition of a single-valued 
polylogarithm on $\IC\backslash\{0,1\}$ is not unique for weight greater than 
two.\footnote{A discussion of the general properties of such functions and the 
relations between various equivalent definitions may be found, for example,  
in~\cite{brownCRAS}. This paper provides a canonical choice of
single-valued multiple polylogarithms.} Different definitions, when evaluated at 1,
gives rise to  a different rational linear combination of  
single-valued multiple zeta. The functions $D_{a,b}(z)$ in~\eqref{e:Dabx} 
give  particular examples of single-valued multiple polylogarithms.

\subsection{The basis of single-valued multiple zetas} 

The dimension $d_w$ of the subspace of multiple zeta values of weight $w$ in the ring 
over $\mathbb Q$ of multiple zeta values is conjecturally given by the expansion of a rational function~\cite{ZagierMZV,Broadhurst:1996kc,Blumlein:2009cf} 
\begin{equation}
\sum_{w=0}^\infty   d_w \, x^w =  {1 \over1-x^2-x^3}\,.  
\end{equation}
The dimension $d_w^{sv}$ of the subspace of weight $w$ in the ring over $\IQ$ of single-valued 
multiple-zeta values is smaller than $d_w$~\cite{Brown:2013gia}. For example,  at weight 11 the  
basis of ordinary multiple zeta values has dimension 9 and  is composed of 
\begin{eqnarray}
&&  \zeta(3,5,3),\, \zeta(3,5)\zeta(3), \, \zeta(3)^2\zeta(5), \, \zeta(11),\cr\cr
&& \zeta(2) \zeta(3)^3,\, \zeta(2)^4\zeta(3),  \,  \zeta(2)^3 \zeta(5),\, 
   \zeta(2)^2 \zeta(7), \, \zeta(2)\zeta(9)\,.
\end{eqnarray}
Since by definition we have  $\zeta_{sv}(2)=0$ all the basis elements
for multiple zeta values on the second line 
are (conjecturally) mapped to zero as elements of single-valued multiple zetas. At weight 11 a further reduction happens since according to~\cite{Brown:2013gia} we have the relation
\begin{equation}
\zeta_{sv}(3,5,3)=2\zeta(3,5,3)-2\zeta(3)\zeta(3,5)-10\zeta(3)^2\zeta(5)\,,
\end{equation}
and since $\zeta_{sv}(3,5)=-10\zeta(3)\zeta(5)$,  the basis of single-valued multiple zetas at weight 11 has dimension 3 and is composed of
\begin{equation}
    \zeta_{sv}(3,5,3),\,  \zeta_{sv}(3)^2\zeta_{sv}(5), \, \zeta_{sv}(11)\,.
\end{equation}
Corrollary~7.4 of~\cite{Brown:2013gia} gives the dimensions $d_w ^{sv}$ of the subspace of  single-valued multiple zetas of weight $w$
\begin{equation}
  \sum_{w=0} ^\infty d^{sv}_w \, x^w = \prod_{n=0}^\infty (1-x^{2n+1})^{-\ell_{2n+1}}  \,,
\end{equation}
where the positive integers $\ell_n$ are defined by the following product relation, 
\begin{equation}
  \prod_{n=1} ^\infty  (1-x^n)^{\ell_n}=1-x^2-x^3   \,.
\end{equation}
Note that, up to weight 12,  the non-zero values for $\ell$ are $\ell_2= \ell_3=\ell_5=\ell_7=\ell_8=\ell_9=\ell_{10}=1$ and $ \ell_{11}=\ell_{12}=2$, so that the non-zero values for $d^{sv}_w$ are  $d_3^{sv}=d_5^{sv}=d_6^{sv}=d_7^{sv}=d_8^{sv}=1$, while $d_9^{sv}=d_{10}^{sv}=2$ and $d_{11}^{sv}=3$ in agreement with our earlier calculation.
Single-valued multiple zeta values are very efficiently obtained
using Schnetz' {\tt Maple}
routines~\cite{SchnetzProc}.\footnote{The procedure in ~\cite{SchnetzProc}
  presents a  Feynman diagram description of single-valued multiple
  polylogarithms reminiscent of the description to be presented in
  this paper.  However, the discussion there is in the context of
  quantum field theories that are quite different from the
  two-dimensional conformal field theory that we are considering, and
  refers to genus-zero functions, which need special treatment.}

\smallskip

We will give an example in section~\ref{sec:three-vert-graphs} of how such single-valued multiple zetas enter into the Laurent expansions of  the modular graph  functions.   They have already appeared in Feynman diagrams in quantum field theory~\cite{Schnetz:2013hqa} and in the low energy expansion of closed string tree level amplitudes~\cite{Stieberger:2013wea,Stieberger:2014hba}.

 \subsection{Single-valued elliptic polylogarithms}
\label{sec:ellpoly}

In the main part of this paper we will consider generalisations of polylogarithms and multiple polylogarithms that are defined  on a general elliptic curve,  that enters, for example,  into the integrand of the genus one closed string amplitude.  An elliptic curve $\mathcal E$ can be described either as 
$\mathcal E\simeq\IC/(\ZZ+\tau \ZZ)$ where $\tau$ is the period ratio or  as 
$\mathcal E\simeq \IC^\times/q^\ZZ$ where $q=\exp(2i\pi \tau)$,  and
$\IC^\times$ is the multiplicative group of non-zero complex numbers.
A point $P$ on the elliptic curve is represented as $\zeta (P)=\exp(2i
\pi (u\tau+ v)) \in \IC^\times/q^\ZZ$ where $u$ and $v$ are cartesian
coordinates in $[0,1]$.  The real part of the period ratio is denoted by
$\tau_1=\Ree(\tau)$ and its imaginary part by $\tau_2=\Imm(\tau)$.

\smallskip

We will make extensive use of the elliptic polylogarithms that were constructed by Zagier~\cite{ZagierDm}.  These can be expressed as infinite sums over positive integers $n$
of images of the polylogarithms $ \Li_k ( z)$
under the translations $z \to z + n \tau$.   To be explicit, these funtions are defined by
\begin{equation}\label{e:Dabq}
  D_{a,b}(q; \zeta)= \sum_{n\geq0}   D_{a,b}(q^n\zeta)+ (-1)^{a+b}\,
  \sum_{n\geq1} D_{a,b}(q^n/\zeta)+{(4\pi \tau_2)^{a+b-1}\over (a+b)!} B_{a+b}(u)\,,
\end{equation}
where the function $D_{a,b}(x)$ was defined in (\ref{e:Dabx}), $B_n$ is the $n$-th Bernoulli polynomial, and $\zeta = e^{2 \pi i (u \tau + v)}$ corresponds to a point on the torus with modulus $q=e^{2 \pi i \tau}$ for $u,v \in [0,1]$.  Zagier  showed that if  $a$ and $b$ are two positive integers, this function is equal to the double sum
\begin{equation}
\label{e:dabdef}
D_{a,b}(q;\zeta) = {(2i\tau_2)^{a+b-1}\over2i\pi}\sum_{(m,n)\neq(0,0)}    {e^{2i\pi(n u-m
      v)}\over (m\tau+n)^a \, (m\bar \tau+n)^b}\,.
\end{equation}
The function $D_{a,b}(q;\zeta)$ is invariant under $\zeta\to q\zeta$ since the expression 
in~\eqref{e:dabdef} is invariant under the shifts $u\to u+1$ and $v\to v+1$, therefore this 
is a function on the elliptic curve. Furthermore,  under the action of $SL(2,\ZZ)$    
\begin{eqnarray}
\label{mod}
  \tau &\to& {\alpha \tau+\beta \over \gamma \tau+\delta }\qquad  \alpha \delta - \beta \gamma =1, \qquad \alpha, \beta, \gamma, \delta \in\ZZ\nn\\
(u,v)&\to&  (\delta u- \gamma v,- \beta u + \alpha v)
\end{eqnarray}
 $D_{a,b}(q;\zeta)$ transforms as a modular form of weight $(1-b,1-a)$, namely, 
\bea
D_{a,b}(q;\zeta)  \, \to \, \left(\gamma \tau +\delta  \right)^{1-b}\left( \gamma \bar\tau +\delta \right)^{1-a}\, D_{a,b}(q;\zeta)\,.
\label{Dabtrans}
\eea

\smallskip

The special case $a=b$ will be of particular importance in this paper and is given by
\begin{equation}\label{e:DaaqDef}
  D_{a,a}(q; \zeta)= \sum_{n\geq0}   D_{a,a}(q^n\zeta)+ 
  \sum_{n\geq1} D_{a,a}(q^n/\zeta) + {(4\pi \tau_2)^{2a-1}\over (2a)!} B_{2a}(u)\,,
\end{equation}
where $D_{a,a}(x)$ is given by \eqref{e:Daax}. The elliptic polylogarithms defined in this manner are manifestly single-valued, as follows from their form in~\eqref{e:dabdef}.

 \subsection{Elliptic multiple polylogarithms}
\label{sec:ellmultpoly}

Elliptical generalisations of  multiple polylogarithms have been defined in~\cite{brownlevin}.  These are holomorphic  functions of  $q$.  These are of relevance to loop amplitudes in open string theory~\cite{Broedel:2014vla} as well as arising   in dimensionally regularised Feynman graphs in quantum field theory at two loops~\cite{Bloch:2013tra,Adams:2014vja,Adams:2015ydq} and three loops~\cite{Bloch:2014qca}.  References~\cite{Bloch:2013tra,Bloch:2014qca}  give a motivic approach to these objects.
We will not review the detailed form of these functions since they do not arise in the rest of this paper.

  \smallskip
 
It will be established later in this paper that we may define {\it single-valued elliptic multiple polylogarithms}  as integrals of products of $D_{1,1}(q;\zeta)$ factors over the torus, with one argument left unintegrated.  Modular graph  functions  may then be described as the values of single-valued elliptic multiple polylogarithms with the unintegrated argument set equal to~1.  The terminology {\it single-valued} arises because single-valued elliptic multiple polylogarithms  have no monodromy as  functions of their unintegrated arguments.  This mimics the terminology used for single-valued multiple zeta values, which are the values of single-valued multiple polylogarithms with arguments set equal to~1.  

\smallskip 

The single-valued  functions that we will describe should be distinguished from the elliptic multiple polylogarithms that arise in~\cite{brownlevin} (mentioned above), which are holomorphic functions of  $q$.    It would be very interesting to discover a precise relationship between the single-valued and non-single-valued elliptical multiple polylogarithms.

\bigskip

%%%%%%%%%%%%%%%%%%%%%%%%%%%%%%%%%%%%%%%%%%%%%
%%%%%%%%%%%%%%%%%%%%%%%%%%%%%%%%%%%%%%%%%%%%%
\section{Feynman graphs associated with a torus}
\label{sec:string}
%%%%%%%%%%%%%%%%%%%%%%%%%%%%%%%%%%%%%%%%%%%%%
%%%%%%%%%%%%%%%%%%%%%%%%%%%%%%%%%%%%%%%%%%%%%

Our motivation for considering  modular graph functions originated from the study of the low energy 
expansion of the genus-one  four-graviton   amplitude in Type II superstring theory.  The graphs 
generated in this expansion  have up to four vertices joined by the scalar Green function on the torus, 
as will be reviewed below. However, the structure of such graphs generalises in an obvious fashion to 
graphs with an arbitrary number of vertices, and it is these general graphs that will be considered in this 
section and sections~\ref{sec:graphs},~\ref{sec:svEgraphs}, and~\ref{sec:6} below. Actually, this obvious generalisation 
will not quite encompass all the contributions that can arise in the $N$-string amplitudes at genus-one 
for $N>4$, as was discussed to a limited extent in \cite{Green:2013bza}.  We will comment on the 
structure of such graphs in section~\ref{sec:highern}.

\smallskip

The four-graviton amplitude at genus one is given by an integral over
the moduli space $\cM_1$ of a torus $\Sigma$ of a partial amplitude
$\cB_4$ evaluated at fixed modulus.  A torus $\Sigma$ with modulus $\tau$ may be represented
in the complex plane  by $\IC/(\ZZ\tau + \ZZ)$ and  parametrised by a
complex coordinate  $z=u \tau + v$ or equivalently by two real
coordinates $u,v \in \IR/\ZZ$. The volume form on $\Sigma$ is
normalised to $d^2 z= i dz \wedge d\bar z /2= - \tau_2 du \wedge
dv$. The moduli space $\cM_1$ of orientable genus-one Riemann surfaces
(or tori) may be represented by a fundamental domain for the action of
$PSL(2,\ZZ)$ on the Poincar\'e upper half plane, which may be
parametrised by $\tau=\tau_1+i \tau_2$ with $\tau_1, \tau_2 \in \IR$
and $\cM_1=\{\tau, ~ 0 < \tau_2, ~ |\tau| > 1, ~ |\tau_1|< \frac12\}$ and the contribution from the boundaries 
$\{\tau, ~ 0 < \tau_2, ~ |\tau| \geq1 , ~ \tau_1=-\frac12\}$ and $\{\tau, ~ 0 < \tau_2, ~ |\tau| = 1, ~ -\frac12<\tau_1\leq 0\}$. We define the following  partial amplitude $\cB_N (s_{ij}| \tau)$ for arbitrary $N$
\begin{equation}
\label{e:B1}
\cB_N (s_{ij} | \tau)
 = 
 \prod _{n=1}^N \int _\Sigma {d^2z_n \over \tau_2} \,
\exp \left( \sum _{1\leq i<j \leq N} s_{ij}  \, G(z_i-z_j |\tau ) \right).
\end{equation}
The scalar Green function $G(z_i-z_j|\tau)$ on the torus will be
discussed in the subsection below. The parameters $s_{ij}$ are related to the momenta $k_i$
of $N$ gravitons,  for $i=1,\dots, N$. Each momentum $k_i$ is a null-vector in 10-dimensional 
Minkowski space-time $\IR^{10}$ so that $k_i \cdot k_i=0$. The relation between the $s_{ij}$ and $k_i$ is 
given by $s_{ij} = -\ell_s^2 \, k_i \cdot k_j/2$, where $\ell_s$ is the string length. The null-vector condition on 
$k_i$ and overall momentum conservation $\sum _{i=1}^N k_i=0$ impose the following relations between 
the parameters $s_{ij}$
\bea
s_{ii}=0, \hskip 0.7in \sum _{j=1}^N s_{ij} =0, \hskip 0.4in i=1, \dots, N.
\label{3a1}
\eea
Actually, any partition of the sum over momenta into two disjoint sets will give rise to extra linear {\it partition relations}  between the variables $s_{ij}$. As a result, for $N\leq 3$, we have $s_{ij}=0$ for all $1\leq i,j \leq 3$ and thus $\cB_1=\cB_2=\cB_3=1 $. The first non-trivial case is the four-graviton amplitude with $N=4$,  and extra  {\it partition relations} $s_{12}=s_{34}, \, s_{23}=s_{14}, \, s_{13}=s_{24}=-s_{12}-s_{14}$ resulting from partitioning the momenta into pairs.

\smallskip
 
The partial amplitude $\cB_N(s_{ij} | \tau)$, with the parameters $s_{ij}$ subject to the relations (\ref{3a1}),
is well-defined for any value of $N$, and will be taken here as the
generating function for modular graph functions, although it is only
when $N=4$ that $\cB_N (s_{ij}|\tau)$  corresponds precisely to the
partial closed superstring amplitude.\footnote{See the discussion in
  section~\ref{sec:highern} for some comments concerning the $N>4$
  amplitude. The genus-one four-graviton amplitude is given by
\begin{equation}
\mathcal A(\epsilon_i,k_i) = 2\pi \,\kappa_{10}^2 \mathcal R^4\, \mathcal B_4(s_{12},s_{13})  
\end{equation}
where $\kappa_{10}^2$ is the ten-dimensional  Newton constant 
and the graviton polarisations, which
appear in the linearised curvature tensor $\mathcal R$,   are denoted by $\epsilon_i$.
The $\alpha'$ expansion of  $\mathcal
B_4(s_{12},s_{13})$ only involves contributions from
products of the Green functions connecting at most $N=4$ vertices on
the world-sheet.
} The restrictions imposed by the extra {\it partition relations} between the parameters $s_{ij}$ will be left unenforced, as  they may be imposed easily, if needed, when constructing the partial amplitudes  for actual string processes.

\subsection{The scalar Green function}

A scalar Green function $G(z|\tau)$ is defined to be the inverse of
the Laplace operator $\Delta= 4 \p_{\bar z} \p_z$ on the torus,
$\Sigma$. Since $\cB_N$ is invariant under a $z$-independent shift of
the Green function $G(z|\tau)$ because of momentum conservation,  we may  restrict  the Green function to have vanishing integral on $\Sigma$. Therefore, $G(z|\tau)$  obeys the following equations
\bea
\label{3b0}
\Delta G(z|\tau) = - 4 \pi \delta ^{(2)} (z) + { 4 \pi \over \tau_2} \hskip 1in \int _\Sigma d^2z \, G(z|\tau)=0\, .
\eea
This determines the modular invariant formula for $G$  in terms of the Jacobi $\vartheta_1$-function and the Dedekind $\eta$-function
\bea
G(z|\tau) = - \ln \left | { \vartheta _1 (z |\tau) \over \eta (\tau) } \right |^2 - { \pi \over 2 \tau_2 } (z-\bar z)^2\, .
\eea
Equivalently, the Green function may be expressed as a Fourier sum in terms of the real variables $u,v \in \IR/\ZZ$ defined by $z=u\tau +v$
\bea
\label{3b1}
G(z|\tau) =  \sum_{(m,n)\neq(0,0)}
  {\tau_2 \over \pi |m\tau+n|^2} \, e^{2i\pi (m v-nu)}\,.
\eea
 The integers $m,n$  parametrise \emph{ the discrete momenta } $p=m\tau +n \in \ZZ \tau + \ZZ$ of the Fourier modes on the torus.  

\smallskip

To make contact with polylogarithms, we change variables for both the modulus $\tau$ and the coordinate $z$ on the torus, to new variables defined in terms of $\tau$ and $z$  by
\bea
q= e^{ 2 \pi i \tau} 
\hskip 1in \zeta = e^{2 \pi i z} = e^{2 \pi i (u\tau + v)}
\eea
The condition $0 <\tau_2$ guarantees that $|q|<1$, while periodicity of $\zeta$ in $v$  shows that $\Sigma$ is parametrised by $\zeta \in \IC^\times/q^\ZZ$. Comparing the expression for $D_{a,b}(q;\zeta)$ of (\ref{e:dabdef}) for $a=b=1$ with the Fourier sum in  (\ref{3b1}), we are led to the following identification \cite{ZagierDm,Green:1999pv} 
\bea
\label{e:D11def1}
 D_{1,1}(q; \zeta)  = G(z|\tau) \,.
\eea
Similarly, the expression $D_{1,1}(q; \zeta_i/\zeta_j)$ will equal the Green function $G(z_i-z_j|\tau)$ on the torus between the points $i$ and  $j$, with $\zeta_i/\zeta_j =e^{2i\pi( z_i-z_j)}$. In view of this identity, we shall often refer to $D_{1,1}(q;\zeta)$ as the Green function as well.

\smallskip

Importantly for the subsequent discussions, the Green function
$D_{1,1}(q;\zeta)$  admits a simple expression in terms of
polylogarithms $2\Ree \Li_1(\cdot)$, which are single-valued on
$\IC\backslash\{0,1\}$,  evaluated on an elliptic curve
\begin{equation}
\label{e:D11Li1}
  D_{1,1}(q;\zeta)=2\Ree\left( \sum_{n\geq0}
    \Li_1(q^n\zeta)+\sum_{n\geq1} \Li_1(q^n/\zeta)+
    \pi \tau_2 B_2(u)  \right)\,,
\end{equation}
where $\zeta=q^u \, e^{2i\pi v}$ with $u,v\in[0,1]$, and $B_2(u)$ is the second Bernoulli polynomial.

\subsection{Modular graph functions}

The Fourier representation of $G(z|\tau)$ makes it transparent   that the Green function $G(z|\tau)$ is invariant under the modular transformations of (\ref{mod}). As a result, the function $\cB_N (s_{ij} |\tau)$ is also modular invariant when the variables $s_{ij}$ are left unchanged under the modular transformation. 
The series expansion of $\cB_N (s_{ij} |\tau)$ in powers of $s_{ij}$ has a finite radius of convergence,
and may be organised graphically in what physicists refer to as Feynman graphs, in this case for a conformal scalar  field on the torus. Each term in this expansion corresponds to a specific graph, and will evaluate to a 
specific non-holomorphic modular function, whence the terminology {\it modular graph function}.

\smallskip

The building block of the modular graphs is the scalar Green function $G(z_i-z_j|\tau)$, which we associate 
with an edge in the graph between vertices $z_i$ and $z_j$. Alternatively, and more conveniently for 
application to polylogarithms, the scalar Green function may be set to the function 
$D_{1,1}(q;\zeta_i/\zeta_j)$. We have the following graphical representation
\begin{center}
\tikzpicture[scale=1.7]
\scope[xshift=-5cm,yshift=-0.4cm]
\draw (1,0) -- (2.5,0) ;
\draw (1,0) [fill=white] circle(0.05cm) ;
\draw (2.5,0) [fill=white] circle(0.05cm) ;
\draw(4.5,0) node{$= \ D_{1,1} (q;\zeta_i/\zeta_j)= G(z_i-z_j|\tau) $.};
\draw (1,0.3) node{$\zeta_i$};
\draw (2.5,0.3) node{$\zeta_j$};
\endscope
\endtikzpicture
\end{center}
The Green function between two vertices $\zeta_i$ and $\zeta_j$ may enter into a given graph raised to an integer power $n$, for which we shall use the following graphical representation, 
\begin{center}
\tikzpicture[scale=1.7]
\scope[xshift=-5cm,yshift=-0.4cm]
\draw (1,0) -- (2.5,0) ;
\draw (1.75,-0) [fill=white] circle(0.15cm) ;
\draw (1,0)    [fill=white] circle(0.05cm) ;
\draw (2.5,0) [fill=white] circle(0.05cm) ;

\draw (1.75,0.0) node{$n$};
\draw(4.6,0) node{$~ = \ D_{1,1} (q;\zeta_i/\zeta_j)^n= G(z_i-z_j|\tau)^n $.};
\draw (1,0.3) node{$\zeta_i$};
\draw (2.5,0.3) node{$\zeta_j$};
\endscope
\endtikzpicture
\end{center}
Finally, the integration at one vertex over a product of Green functions ending at that vertex will be denoted by a filled black dot, in contrast with an unintegrated vertex which will be represented by an unfilled dot. The basic ingredient in the graphical notation is depicted in the graph below

\smallskip

\begin{center}
\tikzpicture[scale=1.7]
\scope[xshift=-5cm,yshift=-0.4cm]
\draw (2,0.7) -- (1,0) ;
\draw (2,0.7) -- (1.5,0) ;
\draw (2,0.7) -- (2.5,0) ;
\draw (2,0.7) -- (3,0) ;
\draw (2,0.2) node{$\cdots$};
\draw [fill=black]  (2,0.7)  circle [radius=.05] ;
\draw (1,0)    [fill=white] circle(0.05cm) ;
\draw (1.5,0) [fill=white] circle(0.05cm) ;
\draw (2.5,0)    [fill=white] circle(0.05cm) ;
\draw (3,0) [fill=white] circle(0.05cm) ;
\draw(5.2,0) node{$\displaystyle= \ \int _\Sigma {d^2 \log \zeta \over 4 \pi ^2 \tau_2} \prod _{i=1}^r D_{1,1} (q;\zeta/\zeta_i)$.};
\draw (1,-0.3) node{$\zeta_1$};
\draw (1.5,-0.3) node{$\zeta_2$};
\draw (2.57,-0.3) node{$\zeta_{r-1}$};
\draw (3.1,-0.3) node{$\zeta_r$};
\endscope
\endtikzpicture
\end{center}

\smallskip

The low energy expansion of $\cB_N$ is obtained simply by expanding the exponential in~\eqref{e:B1} in powers of $s_{ij}$, or equivalently by expanding in powers of $\ell_s^2$, while keeping the momenta $k_i$ fixed.  The coefficient of the term which is homogeneous of degree $w$ in the $s_{ij}$ variables  is given by a sum over all possible graphs with $w$ Green functions  joining at most $N$ vertices, as  studied for the special case of $N=4$ in~\cite{Green:1999pv,Green:2008uj,D'Hoker:2015foa,Basu:2015ayg}.  Each weight $w$ graph $\Gamma_{w,N}$ is evaluated  by the integration of the product of Green functions over the position of the vertices $\zeta_i$ on the torus $\Sigma$. When no confusion is expected to arise, we shall abbreviate $\Gamma _{w,N}$ by simply $\Gamma$. The resulting function will be denoted $I_\Gamma (q)$ and is given by
\begin{equation}
\label{e:IGammaDef}
  I_\Gamma (q)=   \prod_{k=1}^N \int _\Sigma   {d^2\log\zeta_k\over 4\pi^2 \tau_2}  \,
  \prod_{1\leq i<j\leq N} D_{1,1}(q;\zeta_i/\zeta_j)^{n_{ij}}\, ,
\end{equation}
where $N$ is the number of vertices and $n_{ij}$ is the number of
Green functions joining the vertices $i$ and $j$.  The numbers
$n_{ij}$ are the entries of the adjacency matrix of the graph $\Gamma
= \Gamma_{w,N}$.  By construction, the integral $I_\Gamma (q)$ is
modular invariant, as it arises from the expansion in powers of
$s_{ij}$ of the modular invariant generating function
$\cB_N(s_{ij}|\tau)$, and therefore associates with a graph $\Gamma$ a
non-holomorphic modular function $I_\Gamma (q) $ of $q$ and $\bar q$.

\smallskip

An alternative evaluation of the modular graph function  $I_\Gamma (q)$ is obtained by using the Fourier representation 
\bea
D_{1,1}(q;\zeta) = \sum_{(m,n) \not= (0,0)}
  {\tau_2 \over \pi |m\tau +n|^2} \, e^{2i\pi (m v-nu)}\,, 
\eea
where $\zeta = e^{2\pi i (u\tau+v)}$. The integration over $\Sigma$ of the position $\zeta_i$ of the vertex $i$ enforces momentum conservation on all the Green functions that enter the vertex $i$. Carrying out the integrations over $\Sigma$ for all the vertex positions $\zeta _i$ for $i=1,\dots, N$ gives the constrained multiple sum representation for the graphs, as was studied already earlier for the case $N=4$ in~\cite{Green:1999pv,Green:2008uj,D'Hoker:2015foa,D'Hoker:2015zfa,Basu:2015ayg}. The general form of the sum may be schematically represented as follows
\begin{equation}
\label{gengraph}
I_\Gamma (q) =  \sum_{p_1,\dots,p_w\in \ZZ \tau + \ZZ}' ~ \prod_{\alpha =1}^w
  {\tau_2\over  \pi|p_\alpha |^2}\, \prod_{i =1}^N
  \delta \left ( \sum_{\alpha =1}^w C_{i \alpha}  p_\alpha \right )\,.
\end{equation}
Here, the prime above the summation symbol indicates that the sums over $p$ exclude the value 0; the Kronecker $\delta$ symbol takes the value 1 when its argument vanishes and zero otherwise; the coefficients  
$C_{i \alpha}$ are given as follows
\bea
C_{i \alpha} = \left \{ 
\begin{matrix} 
\pm 1 & \hbox{if edge $\alpha$ ends on vertex $i$} \\ & \\
0  & \hbox{otherwise} 
\end{matrix} 
\right .
\eea 
the sign being determined by the orientation of the momenta.   Note that, given the weight $w$, all the information on the graph $\Gamma$ is encoded in the coefficients $C_{i \alpha}$, the other parts of the sum in (\ref{gengraph}) being the same for all $w$. This momentum representation of the modular graph function clearly confirms its modular invariance.

\subsection{Single-valued nature of the modular graph functions}\label{sec:3.3}

In order to understand  the  statement that the modular graph functions
are {\it single-valued}, we now show  that   they are naturally
expressed in terms of  elliptic analogues of single-valued multiple
polylogarithms. We will call these functions  {\it single-valued elliptic multiple polylogarithms}. 
We will demonstrate that the function $I_\Gamma(q)$ is a special value
of a more general function, $I_\Gamma (q; \zeta)$, associated with the
graph $\Gamma =\Gamma_{w,N}$ when all but one of the vertices, say
vertex $j$,  are integrated and the lines meeting the unintegrated
vertex $j$ are made to terminate at two separate points, $\zeta_j$ and
$\zeta_j'$. By translation invariance on the torus, we may set $\zeta_j'=1$ without loss of generality. The function $I_\Gamma(q,\zeta)$ has no monodromies in
$\zeta$ and is therefore single-valued on the torus. 

\smallskip

To be as concrete as possible, we shall concentrate here on the case of $N=4$, but our considerations may be easily generalised to other values of $N$.  An arbitrary graph $\Gamma = \Gamma _{w,4}$ of weight $w$ and with four vertices may be labeled as follows
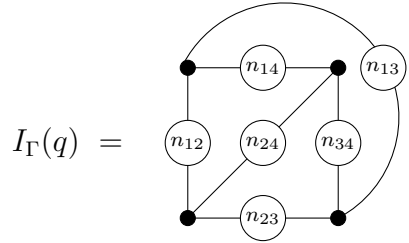
\begin{center}
\tikzpicture[scale=2]
\scope[xshift=-5cm,yshift=-0.4cm]
\draw (0,0) -- (1,0) ;
\draw (0,0) -- (0,1);
\draw (1,1) -- (1,0) ;
\draw (1,1) -- (0,1);
\draw (0,0) -- (1,1) ;
\draw (1,0) arc (-62:152:.76cm);
\draw (0,0.5) [fill=white] circle(0.15cm) ;
\draw (0,0.5)  node{${\scriptstyle n_{12}}$};
\draw (0.5,0) [fill=white] circle(0.15cm) ;
\draw (0.5,0)  node{${\scriptstyle n_{23}}$};
\draw (1,0.5) [fill=white] circle(0.15cm) ;
\draw (1,0.5)  node{${\scriptstyle n_{34}}$};
\draw (0.5,1) [fill=white] circle(0.15cm) ;
\draw (0.5,1)  node{${\scriptstyle n_{14}}$};
\draw (0.5,0.5) [fill=white] circle(0.15cm) ;
\draw (0.5,0.5)  node{${\scriptstyle n_{24}}$};
\draw (1.3,1) [fill=white] circle(0.15cm) ;
\draw (1.3,1)  node{${\scriptstyle n_{13}}$};
\draw [fill=black]  (0,0)  circle [radius=.05] ;
\draw [fill=black] (1,0)  circle [radius=.05] ;
\draw [fill=black] (0,1)  circle [radius=.05] ;
\draw [fill=black] (1,1)  circle [radius=.05] ;
\draw (-0.8,0.5) node{$I_\Gamma (q) ~ =$};
\endscope
\endtikzpicture
\end{center}
 The resulting elliptic function $I_\Gamma (q; \zeta_1)$  is illustrated by the graph below with five vertices, of which two are unintegrated. The unintegrated vertex corresponds here to the vertex $j=1$ in the figure
 \begin{center}
\tikzpicture[scale=2]
\scope[xshift=-5cm,yshift=-0.4cm]
\draw (0,0) -- (1,0) ;
\draw (0,0) -- (0,1);
\draw (1,1) -- (1,0) ;n
\draw (1,1) -- (0,1);
\draw (0,0) -- (1,1) ;
\draw (1,0) -- (2,0);
%\draw (1,0) to[out=0,in=-90] (1.5,1) to[out=90,in=45] (.5,1.5) to (0,1);
%\draw (1,0) arc (-90:180:1cm);
\draw (0,0.5) [fill=white] circle(0.15cm) ;
\draw (0,0.5)  node{${\scriptstyle {n_{12}}}$};
\draw (0.5,0) [fill=white] circle(0.15cm) ;
\draw (0.5,0)  node{${\scriptstyle n_{23}}$};
\draw (1,0.5) [fill=white] circle(0.15cm) ;
\draw (1,0.5)  node{${\scriptstyle n_{34}}$};
\draw (0.5,1) [fill=white] circle(0.15cm) ;
\draw (0.5,1)  node{${\scriptstyle n_{14}}$};
\draw (0.5,0.5) [fill=white] circle(0.15cm) ;
\draw (0.5,0.5)  node{${\scriptstyle n_{24}}$};
\draw (1.5,0) [fill=white] circle(0.15cm) ;
\draw (1.5,0)  node{${\scriptstyle n_{13}}$};
\draw [fill=black]  (0,0)  circle [radius=.05] ;
\draw [fill=black] (1,0)  circle [radius=.05] ;
\draw [fill=white] (0,1)  circle [radius=.05] ;
\draw [fill=black] (1,1)  circle [radius=.05] ;
\draw [fill=white] (2,0)  circle [radius=.05] ;
\draw (0,1.3) node{$\zeta_1$};
\draw (2,0.3) node{$\zeta_1' $};
\draw (-1,0.5) node{$I_\Gamma (q;\zeta_1/\zeta_1') ~ =$};
\endscope
\endtikzpicture
\end{center} 
The  black dots  represent  integrated vertices, while the white dots continue to represent unintegrated vertices. To such a graph we associate an elliptic function that results from the integration over the remaining three vertices, which here are $k=2,3,4$
\begin{equation}
\label{e:IGammaDefzeta}
  I _\Gamma (q; \zeta ) = 
 \prod_{k=2}^{4} \int _\Sigma {d^2\log\zeta_k \over 4 \pi^2 \tau_2} \,
  \prod_{1\leq i<j\leq 4} D_{1,1}(q;\zeta_j/\zeta_i)^{n_{ij}} \, 
\left ( { D_{1,1} (q;\zeta_1'/\zeta_3)  \over D_{1,1} (q;\zeta _1/\zeta_3)  }  \right )^{n_{13}} \,.
\end{equation} 
The function $I_\Gamma (q;\zeta)$ depends only on the ratio $\zeta=\zeta_1/\zeta_1'$ 
due to translation invariance on the torus. Since the Green function $D_{1,1}(q;\zeta)$ is 
a single-valued function of $\zeta$ the function  $I_\Gamma (q; \zeta)$  is a single-valued 
function of $\zeta$.   Translation invariance implies that the  full  integral~\eqref{e:IGammaDef}
is recovered by identifying the points $\zeta_1$ and $\zeta_1'$ in the above graph, in other words, by setting $\zeta=1$.
This is easily shown using the representation of ~\eqref{gengraph} in which the discrete momentum is preserved at each vertex.

\smallskip

\subsection{ Proposition}
\label{sec:Prop} {\it Any modular graph function $I_\Gamma (q)$
may be obtained as the value of a single-valued elliptic multiple polylogarithm 
$I_\Gamma (q; \zeta)$ evaluated at the point $\zeta=1$.}

%%%%%%%%%%%%%%%%%%%%%%%%%%%%%%%%%%%%%%%%%%%%%
%%%%%%%%%%%%%%%%%%%%%%%%%%%%%%%%%%%%%%%%%%%%%
\section{Examples of modular graph functions}
\label{sec:graphs}
%%%%%%%%%%%%%%%%%%%%%%%%%%%%%%%%%%%%%%%%%%%%%
%%%%%%%%%%%%%%%%%%%%%%%%%%%%%%%%%%%%%%%%%%%%%

In this section we shall  illustrate the structure of modular graph functions $I_\Gamma (q)$
and their associated single-valued elliptic multiple polylogarithms $I_\Gamma (q;\zeta)$ on
some simple but significant examples. It will be convenient to characterise the families of graphs 
under consideration by the number of loops $L$ of the graph $\Gamma $ in the corresponding 
modular graph function $I_\Gamma (q)$.

\subsection{General one-loop graphs}
\label{sec:aprop}

The first family of examples is based on modular graph functions $I_\Gamma (q)$ for graphs $\Gamma$ 
with a single loop. The associated single-valued elliptic multiple polylogarithm results from a linear chain 
graph in which $a$ Green functions are  concatenated and integrated over their $a-1$ junction points. 
Such linear chain graphs are important in their own right, but will also provide natural building blocks for 
higher graphs that involve linear chains. We introduce the graphic notation given in the figure below
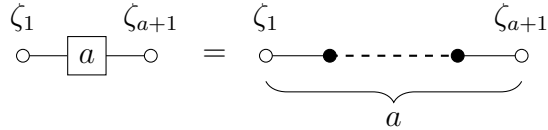
\begin{center}
\tikzpicture[scale=1.7]
\scope[xshift=-5cm,yshift=-0.4cm]
\draw (0,0) -- (1,0) ;
\draw (0.35,-0.15) [fill=white] rectangle (0.65,0.15) ;
\draw (0.5,0) node{$a$};
\draw[fill=white] (0,0)  circle [radius=.05] ;
\draw[fill=white] (1,0)  circle [radius=.05] ;
\draw (1.5,0) node{$=$};
\draw (1.9,0) -- (2.4,0) ;
\draw[dashed, thick] (2.4,0) -- (3.4,0) ;
\draw (3.4,0) -- (3.9,0) ;
\draw[fill=white] (1.9,0)  circle [radius=.05] ;
\draw[fill=black] (2.4,0)  circle [radius=.05] ;
\draw[fill=black] (3.4,0)  circle [radius=.05] ;
\draw[fill=white] (3.9,0)  circle [radius=.05] ;
\draw [decorate,decoration={brace,mirror,amplitude=8pt},xshift=0pt,yshift=0pt]
(1.9,-0.2) -- (3.9,-0.2)  ;
\draw (2.9,-0.5) node{$a$};
\draw (1.9, 0.3) node{$\zeta_1$};
\draw (3.9, 0.3) node{$\zeta_{a+1}$};
\draw (0, 0.3) node{$\zeta_1$};
\draw (1, 0.3) node{$\zeta_{a+1}$};
\endscope
\endtikzpicture
\end{center}
Recall that the black dots correspond to integrated vertices, while the white dots are unintegrated and the corresponding vertices are evaluated at the labels of these points. The weight of the graph is $w=a$, while its number of vertices is $N=a+1$. 
The associated elliptic function may be evaluated by performing the
$a-1$ integrations over the vertices $\zeta _j$ for $j=2,\dots, a$. A
convenient way to carry out this evaluation is with the help of the
Fourier representation of (\ref{e:dabdef}). By comparing the result
with the Fourier representation given in (\ref{e:dabdef}) for $D_{a,a}(q;\zeta)$, we readily find
\begin{equation}
 \prod_{k=2}^{a} \int _\Sigma {d^2\log \zeta_k \over 4 \pi^2 \tau_2}  \,
\prod_{j=1}^{a} D_{1,1}(q;\zeta_{j+1}/\zeta_j) 
  =
(-4\pi \tau_2)^{1-a} D_{a,a}\left(q;\zeta_{a+1}/\zeta_1\right) \,,
\end{equation}
which is the single-valued elliptic polylogarithm  defined in \cite{ZagierDm}.
The single-valuedness in $\zeta= \zeta _{a+1}/\zeta _1$ of $D_{a,a}(q;\zeta)$ is  again a 
consequence of this integral representation, as explained in section~\ref{sec:string}.
We may summarise this results by the graphical representation for the linear chain graph in the figure below
\begin{center}
\tikzpicture[scale=1.7]
\scope[xshift=-5cm,yshift=-0.4cm]
\draw (0,0) -- (1,0) ;
\draw (0.35,-0.15) [fill=white] rectangle (0.65,0.15) ;
\draw (0.5,0) node{$a$};
\draw[fill=white] (0,0)  circle [radius=.05] ;
\draw[fill=white] (1,0)  circle [radius=.05] ;
\draw(2.6,0) node{\qquad $= \, ( - 4 \pi \tau_2)^{1-a}  D_{a,a} \left(q; \zeta_{a+1}/\zeta_1\right)$\, .};
\endscope
\endtikzpicture
\end{center}
By evaluating the open linear chain graph when the vertices $\zeta _1$ and $\zeta _{a+1}$ coincide yields a one-loop chain graph, depicted in the figure  below
\begin{center}
\tikzpicture[scale=1.3]
\scope[xshift=-5cm,yshift=-0.4cm]
\draw [draw=black]       (0,0) circle (1 and .5);
\draw[fill=black] (1,0)  circle [radius=.05] ;
\draw (-0.8,-0.2) [fill=white] rectangle (-1.2,0.2) ;
\draw (-1,0) node{$a$};
\draw (3,0) node{$= \, (-4 \pi \tau_2)^{1-a} D_{a,a}(q;1)$\, .};
\endscope
\endtikzpicture
\end{center}
Upon setting $\zeta=1$ in the argument of $D_{a,a}(q;\zeta)$ in~\eqref{e:dabdef}   we see that
\begin{equation}
\label{e:eisengraph}
(-4 \pi \tau_2)^{1-a}\,D_{a,a}(q;1)=E_a(q)
=\sum_{(m,n)\neq(0,0)} {\tau_2^a\over \pi ^a  |m\tau+n|^{2a}}\,.
  \end{equation}
The representation of the real analytic $E_a$ Eisenstein series as a one-loop graph was
used in~\cite{Green:1999pv,Green:2008uj,D'Hoker:2015foa}. 
 The Eisenstein series is the special value at $\zeta=1$ of the single-valued elliptic polylogarithm function
$D_{a,a}(q;\zeta)$.

\smallskip

%---------------------------------------------------------------------------------
\subsection{General two-loop graphs}

The general two-loop graph that produces a modular graph function is given by the modular functions 
$I_\Gamma (q)= C_{a,b,c}(q)$ introduced and studied extensively  in
\cite{D'Hoker:2015foa}.\footnote{This is a slight change of notation from \cite{D'Hoker:2015foa} where we used
  $C_{a,b,c}(\tau)$ to represent this function.} 
These graphs always have two trivalent vertices, along with $a+b+c-3$ bivalent vertices. For any 
assignment of value of $a,b,c \geq 1$, a single-valued elliptic multiple polylogarithm 
$I_\Gamma (q;\zeta) =C_{a,b,c}(q;\zeta)$ may be 
constructed from $\Gamma$ by detaching one edge from one of the two trivalent vertices. 
By permutation symmetry of the indices $a,b,c$, we may detach the edge labeled $a$ without loss of generality,
to obtain the graph shown in the figure below
\begin{center}
\tikzpicture[scale=1.5]
\scope[xshift=-5cm,yshift=-0.4cm]
\draw [draw=black, ]        (0,0) circle (1 and .5);
\draw (-2,0) -- (-1,0) ;
\draw (-1.65,-0.15) [fill=white] rectangle (-1.35,0.15) ;
\draw (-0.15,-.65) [fill=white] rectangle (0.15,-.35) ;
\draw (-0.15,0.35) [fill=white] rectangle (0.15,0.65) ;
\draw [fill=black] (-1,0) circle [radius=.05];
\draw [fill=white]  (-2,0) circle [radius=.05] ;
\draw [fill=white]  (1,0) circle [radius=.05] ;
\draw (-2,0.3) node{$\zeta_1$};
\draw (1.2, 0.3) node{$\zeta_1'$};
\draw (-1.5,0) node{$a$};
\draw (0, .5) node{$b$};
\draw (0, -.5) node{$c$};
\draw (2.6,0) node{$~ = ~ C_{a,b,c}(q;\zeta_1/\zeta_1')$ .};
\endscope
\endtikzpicture
\end{center} 
The associated  single-valued elliptic multiple polylogarithm, $C_{a,b,c}(q;\zeta)$ is a function of the ratio $\zeta=\zeta_1/\zeta_1'$ and has the form
\begin{equation}
C_{a,b,c}(q;\zeta)= (-4\pi\tau_2)^{3-a-b-c} \,\int_{\Sigma }\, {d^2\log \zeta_2\over 4\pi^2 \tau_2}\, 
D_{a,a}(q;\zeta/\zeta_2)D_{b,b}(q;\zeta_2)D_{c,c}(q;\zeta_2)    \,.
\end{equation}
 Setting $\zeta=1$ leads to the modular graph function $C_{a,b,c}(q)$  
\begin{equation}
  C_{a,b,c}(q)= C_{a,b,c}(q;1) \, ,
\end{equation}
associated with the Feynman graph
\begin{center}
\tikzpicture[scale=1.5]
\scope[xshift=-5cm,yshift=-0.4cm]
\draw [draw=black, ]        (0,0) circle (1 and .5);
\draw (-1,0) -- (1,0) ;
\draw (-0.15,-0.15) [fill=white] rectangle (0.15,0.15) ;
\draw (-0.15,-.65) [fill=white] rectangle (0.15,-.35) ;
\draw (-0.15,0.35) [fill=white] rectangle (0.15,0.65) ;
\draw [fill=black] (-1,0) circle [radius=.05];
%\draw [fill=black]  (-2,0) circle [radius=.05] ;
\draw [fill=black]  (1,0) circle [radius=.05] ;
\draw (0,0) node{$b$};
\draw (0, .5) node{$a$};
\draw (0, -.5) node{$c$};
\draw (2.4,0) node{$= ~ C_{a,b,c}(q)$ .};
\endscope
\endtikzpicture
\end{center} 
It follows that $C_{a,b,c}(q)$ is the value at $\zeta=1$ of
$C_{a,b,c}(q;\zeta)$, which is a  single-valued function of $\zeta$.  
This illustrates the
general point made in section~\ref{sec:3.3} and Proposition~\ref{sec:Prop}. 

\smallskip

Using the lattice sum expression for $D_{a,a}(q;\zeta)$
given in~\eqref{e:dabdef} leads to 
the multiple sum representation studied in~\cite{D'Hoker:2015foa,D'Hoker:2015zfa}
\begin{equation}
  \label{e:CabcDef}
  C_{a,b,c}(q)=\sum_{p_1,p_2,p_3\in \mathbb  Z\tau+\mathbb Z} ' \,
  \left(\tau_2\over \pi\right)^{a+b+c}  \,
  {\delta(p_1+p_2+p_3) \over |p_1|^{2a} |p_2|^{2b} |p_3|^{2c}}\,,
\end{equation}
where $p_i = m_i \tau + n_i$ parametrises the momenta on the torus $\Sigma$.  
In \cite{D'Hoker:2015foa} linear combinations of the functions $ C_{a,b,c}(q)$ were shown to satisfy Laplace eigenvalue equations with source terms that are quadratic in real analytic Eisenstein series.  In principle, these equations determine the form of these functions. 

\smallskip

When at least one of the indices $a,b,c$ is greater than one, the graph $\Gamma$ contains bivalent vertices in addition to its two trivalent vertices. When this is the case, one may detach the edges meeting at any of the bivalent vertices, to obtain another single-valued elliptic multiple polylogarithm, specified graphically by the figure below
\begin{center}
\tikzpicture[scale=1.5]
\scope[xshift=-5cm,yshift=-0.4cm]
\draw [draw=black, ]        (0,0) circle (1 and .5);
\draw (-2,0) -- (-1,0) ;
\draw (1,0) -- (2,0) ;
\draw (-1.65,-0.15) [fill=white] rectangle (-1.35,0.15) ;
\draw (1.65,-0.15) [fill=white] rectangle (1.35,0.15) ;
\draw (-0.15,-.65) [fill=white] rectangle (0.15,-.35) ;
\draw (-0.15,0.35) [fill=white] rectangle (0.15,0.65) ;
\draw [fill=black] (-1,0) circle [radius=.05];
\draw [fill=black]  (1,0) circle [radius=.05] ;
\draw [fill=white]  (-2,0) circle [radius=.05] ;
\draw [fill=white]  (2,0) circle [radius=.05] ;
\draw (-2,0.3) node{$\zeta_1$};
\draw (2.05, 0.3) node{$\zeta_1'$};
\draw (-1.48,0) node{$a_1$};
\draw (1.51,0) node{$a_2$};
\draw (0, .5) node{$b$};
\draw (0, -.5) node{$c$};
\draw (3.5,0) node{$~ = ~ C_{a_1;b,c;a_2}  (q;\zeta_1/\zeta_1')$  ,};
\endscope
\endtikzpicture
\end{center} 
where $a_1+a_2=a$ and $a_1, a_2 \geq 0$. This case provides the simplest example of a general phenomenon, namely that to a given modular graph function, there will correspond a number of different and generally inequivalent single-valued elliptic multiple polylogarithm functions. Finally, we note that the above modular two-loop graph functions produce, upon detaching both edges on a single trivalent vertex, a star graph with three branches, and thereby a two-variable single-valued elliptic multiple polylogarithm. We shall examine such graphs further in section~\ref{sec:star} below.

\smallskip

 %-------------------------------------------------------------------------
\subsection{General three-vertex graphs}
\label{sec:three-vert-graphs}

The modular graph function associated with an arbitrary  graph with a total of three vertices  is obtained  as a special value of the single-valued elliptic multiple polylogarithm  associated with the following graph with two unintegrated vertices 
\begin{center}
\tikzpicture[scale=1.7]
\scope[xshift=-5cm,yshift=-0.4cm]
\draw (-2,0) -- (-1,0) ;
\draw (-1,0) -- (0,0);
\draw (0,0) -- (1,0);
\draw (-1.5,0) [fill=white] circle(0.15cm) ;
\draw (-1.5,0)  node{$a$};
\draw (-0.5,0) [fill=white] circle(0.15cm) ;
\draw (-0.5,0)  node{$b$};
\draw (.5,0) [fill=white] circle(0.15cm) ;
\draw (.5,0)  node{$c$};
\draw [fill=white]  (-2,0) circle [radius=.05] ;
\draw [fill=black]  (-1,0) circle [radius=.05] ;
\draw [fill=black]  (0,0) circle [radius=.05] ;
\draw [fill=white]  (1,0) circle [radius=.05] ;
\draw (-2,0.3) node{$\zeta_1$};
\draw (1,0.3) node{$\zeta_1'$};
\draw (2.4,0) node{$ = ~ D_{a,b,c}(q; \zeta_1/\zeta_1')$ ,};
\endscope
\endtikzpicture
\end{center}
where the encircled numbers indicate the number of the Green functions $D_{1,1}$ which appear in the corresponding link, following our earlier graphical representations. The corresponding single-valued elliptic multiple polylogarithm evaluates as follows
\bea
D_{a,b,c}(q;\zeta)=  \prod _{k=2,3} \int_\Sigma {d^2\log\zeta_k \over 4\pi^2 \tau_2}  \,
D_{1,1}(q;\zeta/\zeta_2)^a  \, D_{1,1}(q;\zeta_2/\zeta_3)^b \, D_{1,1}(q;\zeta_3)^c   \,,
\eea
which is a single-valued function of $\zeta$. Its restriction to $\zeta=1$ gives
\begin{equation}
  D_{a,b,c}(q)=D_{a,b,c}(q;1)  \,,
\end{equation} 
corresponding to the  modular graph function associated with the  figure below
\medskip
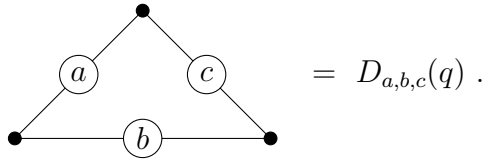
\begin{center}
\tikzpicture[scale=1.7]
\scope[xshift=-5cm,yshift=-0.4cm]
\draw (-1,0) -- (0,1) ;
\draw (-1,0) -- (1,0);
\draw (1,0) -- (0,1);
\draw (-0.5,0.5) [fill=white] circle(0.15cm) ;
\draw (-0.5,0.5)  node{$a$};
\draw (0,0) [fill=white] circle(0.15cm) ;
\draw (0,0)  node{$b$};
\draw (0.5,0.5) [fill=white] circle(0.15cm) ;
\draw (0.5,0.5)  node{$c$};
\draw [fill=black]  (-1,0) circle [radius=.05] ;
\draw [fill=black]  (1,0) circle [radius=.05] ;
\draw [fill=black]  (0,1) circle [radius=.05] ;
\draw (2,.5) node{$= ~D_{a,b,c}(q)$ .};
\endscope
\endtikzpicture
\end{center}

\subsection{General graphs}
The examples of entire classes of modular graph functions, and their associated single-valued elliptic multiple polylogarithms,  provided in the earlier paragraphs lend themselves to a natural generalisation. From a general Feynman graph $\Gamma$, with associated modular graph function $I_\Gamma (q)$, it is possible to construct at least one, but in general several inequivalent, single-valued elliptic multiple polylogarithms $I_\Gamma (q;\zeta)$ by detaching one or several Green function edges from any one of the  vertices in the graph. The original vertex may be placed at an arbitrary reference point, which we choose to be 1, while the vertex of the detached Green function(s) is placed at an arbitrary point $\zeta$. These observations show the validity of Proposition~\ref{sec:Prop} and provide an explicit construction for $I_\Gamma (q;\zeta)$. The significance of the fact that several inequivalent single-valued elliptic multiple polylogarithms arise in this way from a single modular graph function,  remains to be explored in full.

%%%%%%%%%%%%%%%%%%%%%%%%%%%%%%%%%%%%%%%%%%%%%
%%%%%%%%%%%%%%%%%%%%%%%%%%%%%%%%%%%%%%%%%%%%%
\section{Single-valued multiple polylogarithms for graphs}
\label{sec:svEgraphs}
%%%%%%%%%%%%%%%%%%%%%%%%%%%%%%%%%%%%%%%%%%%%%
%%%%%%%%%%%%%%%%%%%%%%%%%%%%%%%%%%%%%%%%%%%%%

In the preceding two sections, we have introduced modular graph functions, and their
relation with single-valued elliptic multiple polylogarithms, leading to Proposition~\ref{sec:Prop}, 
and a series of concrete examples. In the present section, we shall make progress 
towards understanding the description in terms of  (non-single-valued) polylogarithms
of both modular graph functions and single-valued multiple
polylogarithms. 
In particular, we shall advance a {\it \claim}\,  as to this general structure, and offer a 
proof for a certain infinite subclass  of graphs, leaving a complete proof for the general 
case for future investigations.  Assuming the validity of  our conjecture,  we shall infer important 
properties for the ring structure over $\IQ$ of the coefficients of the Laurent 
series in powers of  $\tau_2$ of general modular graph functions in terms of 
single-valued multiple zeta values. The separate terms in the expansions to be discussed later in this section will not be elliptic functions. This is not necessary for the proof of single-valuedness that follows.  However,  the sum of terms is guaranteed to be elliptic since the initial expression for any modular graph function is manifestly elliptic.
The Laurent series is the dominant contribution to these 
modular functions in the limit $\tau_2 \to \infty$.

\smallskip

The main conjecture of this paper applies to any  single-valued multiple polylogarithm $I_\Gamma(q;\zeta)$ associated with a {\sl single-component graph}  modular graph function $I_\Gamma(q)$.
By single-component graph, we mean a connected graph, whose associated modular graph function 
cannot be reduced to the product of lower weight modular graph functions.

%\break

\noindent {\bf \Claim}
%\nobreak
\sm
{\it The single-valued elliptic multiple polylogarithm $I_\Gamma (q;\zeta)$ associated with a single component graph $\Gamma = \Gamma _{w,N}$,  with $w$ Green functions, $N+1$ vertices, and  $L=w-N$ loops, is a linear  combination of  multiple polylogarithms of depth at most  $L$ and weight at most  $w+N-1$.
}

\subsection{Star graphs}\label{sec:star}

We shall now provide a proof of the above \claim\, for the infinite class of {\it star graphs}.
A graph $S$ is a star graph provided it contains a single integrated $n$-valent 
vertex with $n>2$, and an arbitrary number of integrated bi-valent vertices. By construction, 
the graph then has $n$ unintegrated end points $\zeta _i$ with $i=1,\dots, n$. To obtain 
the modular graph function with two $n$-valent vertices associated with  $S$ it 
suffices to set all $\zeta_i$ equal to 1. To obtain single-valued multiple polylogarithms 
from $S$, one may set one non-empty subset of end points to 1 while setting the remaining 
non-empty set of endpoints to $\zeta$. A general  $n$-valent star
graph $S$, with $a_i$ Green functions on leg $i$ of the star, along
with its multi-variable  function value $I_S(q;\zeta_1, \dots, \zeta _n)$,  is  
represented in the figure below

\medskip

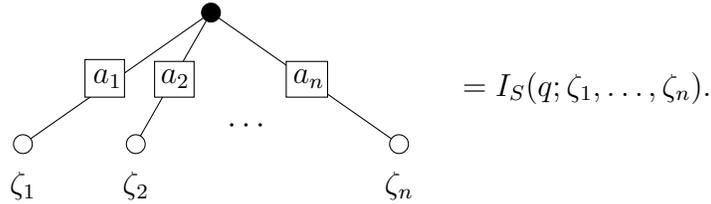
\begin{center}
\tikzpicture[scale=2.5]
\scope[xshift=-5cm,yshift=-0.4cm]
\draw (2,0.7) -- (1,0) ;
\draw (2,0.7) -- (1.6,0) ;
%\draw (2,0.7) -- (2.4,0) ;
\draw (2,0.7) -- (3,0) ;
\draw (2.2,0.1) node{$\cdots$};
\draw [fill=black]  (2,0.7)  circle [radius=.05] ;
\draw (1,0)    [fill=white] circle(0.05cm) ;
\draw (1.6,0) [fill=white] circle(0.05cm) ;
%\draw (2.4,0)    [fill=white] circle(0.05cm) ;
\draw (3,0) [fill=white] circle(0.05cm) ;
\draw (1.33,0.45) [fill=white] rectangle (1.54,0.25) ;
\draw (1.7,0.44) [fill=white] rectangle (1.91,0.25) ;
\draw (2.4,0.44) [fill=white] rectangle (2.62,0.25) ;
\draw(4,0.3) node{$\displaystyle= I_S (q; \zeta _1, \dots, \zeta _n)$.};
\draw (1,-0.22) node{$\zeta_1$};
\draw (1.6,-0.22) node{$\zeta_2$};
%\draw (2.47,-0.3) node{$\zeta_{n-1}$};
\draw (3.0,-0.22) node{$\zeta_n$};
\draw (1.45, 0.35) node{$a_1$};
\draw (1.81, 0.35) node{$a_2$};
\draw (2.52, 0.35) node{$a_n$};
\endscope
\endtikzpicture
\end{center}
Its associated multi-variable modular graph function evaluates to
\bea\label{e:ISdef}
I_S (q; \zeta _1, \dots, \zeta _n)
= \ \int _\Sigma {d^2 \log \zeta \over 4 \pi ^2 \tau_2} 
\prod _{i=1}^n D_{a_i,a_i} (q;\zeta/\zeta_i) \, .
\eea
To prove the \claim\,  for star graphs, we need to show that the result of performing the 
integration of the $n$-valent vertex $\zeta$ over the torus $\Sigma$ produces a linear combination, 
with rational coefficients, of (non-single-valued) multiple polylogarithms, resulting
in a formula schematically represented as follows
\begin{equation}
\label{e:gw}
\sum_{s} \sum_{b_1,\dots,b_s} c_{b_1,\dots,b_s}\,   \Li_{b_1,\dots,b_s}(z_1,\dots,z_s)\,.
\end{equation}
where the coefficients   $c_{b_1,\dots,b_s}$ are rational numbers, the multiple polylogarithm is defined in \eqref{e:Mpl}, and $z_i$ are products  of the arguments, generically of the form
\begin{equation}
  q^m \bar q^{m'} \, \prod_{t}  \zeta_t^{\alpha_t}   \bar\zeta_t^{\beta_t}\,,
\end{equation}
where $\alpha_t$, $\beta_t$ are integers with $1 \leq t \leq s$.
We will proceed by considering some algebraic features of the integrand  
\begin{equation}
\label{e:fDef}
\prod _{i=1}^n D_{a_i,a_i} (q;\zeta/\zeta_i)\,,
\end{equation}
making use of the fact that the generalised Green functions $D_{a_i,
  a_i}(q;\zeta/\zeta_i)$ are themselves polylogarithms of depth 1 and
weight $a_i$  given in~(\ref{e:Dabq}) via~\eqref{e:Daax}, and then
perform the integrations over the phase $v$ and the variable $u$ of
$\zeta = e^{2 \pi i (u \tau + v)}$. The intermediate steps will not be manifestly single-valued in the $\zeta_i$ values but the expression in eq.~\eqref{e:ISdef} guaranties the single-valueness of the final answer.
 
\smallskip

\subsection{Integral over the phase $v$}
\label{sec:intLi}

We begin by recasting the functions $D_{a_i, a_i}(q;\zeta/\zeta_i)$ in
terms of the functions $D_{a_i,a_i}(q^m (\zeta/\zeta_i)^{\pm 1})$ and
their complex conjugate with the help of equation~\eqref{e:DaaqDef},
and then expressing the latter as a linear combination of
polylogarithms ${\rm Li}_k$ using~\eqref{e:Daax}.  In performing the
$v$-integral, the presence of the Bernoulli polynomials $B_{2a}(u)$
 in~\eqref{e:DaaqDef} will be immaterial,  as will be the logarithms
in~\eqref{e:Daax},  since they involve $u$ but do not depend on $v$.  We will suppress their presence in this section. Finally, the arguments entering the polylogarithms ${\rm Li}_k$ in $D_{a_i,a_i}$ depend on $\zeta$ either through a factor linear in $\zeta$, or a factor linear in $\zeta^{-1}$. It will be convenient in the sequel to group together the factors in $\zeta$ and those in $\zeta ^{-1}$, of which we shall assume there are respectively $r$ and $s$. The maximal values of these parameters are characterised by $0 \leq r,s$ and $r+s \leq n$. Thus, the $v$-integrals required to evaluate $I_S$ take the  form
\bea
\label{Jab}
J_{\ba, \bb} (\bx, \by)= \int _0 ^1 dv \, 
\prod _{i =1}^r \Li_{a_i}  (x_i \, e^{2 \pi i v}) \prod _{j=1}^s \Li_{b_j} (y_j \, e^{- 2 \pi i v}) \, ,
\eea
where we have introduced the following notation,
\bea
\ba = (a_1 , \dots, a_r) & \hskip 1in & \bx=(x_1 , \dots, x_r)
\no \\
\bb = (b_1, \dots, b_s) && \by=(y_1, \dots, y_s) \, .
\eea
The entries of the composite  indices $\ba$ and $\bb$ are positive integers, while the entries of $\bx$ and $\by$  depend on $u$, and on positive powers of $q$ and positive powers of $\bar q$, but are independent of $v$. The variables obey $|x_i| , \, |y_j|<1$ for all $i=1,\dots, r$ and $j=1,\dots, s$. In view of these bounds, we may use the series representations for the functions ${\rm Li}$ to perform the integration over $v$ which results in the insertion into the sum of a Kronecker $\delta$-function, 
\bea
J_{\ba, \bb} (\bx, \by) 
=
\sum _{0 < m_1 , \dots , m_r} \, \sum _{0 < n_1, \dots, n_s}
\delta \left ( \sum _{i=1}^r m_i - \sum _{j=1}^s n_j \right )
\prod _{i=1}^r { x_i^{m_i} \over m_i ^{a_i}} 
\prod _{j=1}^s { y_j^{n_j} \over n_j ^{b_j}} \, .
\eea
To prove that this constrained sum may be expressed as a linear combination, with rational coefficients, of functions ${\rm Li}$ evaluated on various combinations of the variables $\bx, \by$, we shall proceed by induction on $s$, at fixed but arbitrary value of $r$. For $r>0$, and $s=0$, the integral vanishes identically.

\sm

To initiate the induction, consider first the case $s=1$. It is straightforward to solve for $n_1$ which is given by
$n_1 = m_1+ \dots + m_r$. The constraint $n_1>0$ imposed by the definition of $J_{\ba,\bb}$  is automatically satisfied since all the integers $m_i$ are strictly positive. Thus, we find
\bea
J_{\ba, b_1} (\bx, y_1) 
=
\sum _{0 < m_1 , \dots , m_r} \, 
{ 1 \over m_1 + \dots + m_r} \,
\prod _{i=1}^r { (x_i y_1) ^{m_i} \over m_i ^{a_i}} \, .
\eea
The summation may be expressed equivalently as the following integral
\bea
J_{\ba, b_1} (\bx, y_1) 
=
\int ^1 _0 {dt \over t}  \, \prod _{i=1} ^r \Li_{a_i} (x_i y_1 t) \, .
\eea
Using the stuffle relations of (\ref{stuffle}), the product over the ${\rm Li}$-functions may be  expressed as a linear combination, with rational coefficients,  of multiple polylogarithms. The one of highest depth $r$ is given by  $\Li _{a_1, \dots, a_r} ( x_1 y_1 t, \dots, x_r y_1 t)$, 
while the one of lowest depth 1 is given by  $\Li_{a_1 + \dots + a_r} ( x_1 \dots x_r y_1 ^r t^r)$, along with all stuffle combinations with rational coefficients and depths in between 1 and $r$. The integral over $t$ of any of these ${\rm Li}$-functions gives back an ${\rm Li}$-function of various combinations of the arguments.
This concludes the proof of the claim for $s=1$.

\sm

Next, we shall assume that the claimed decomposition property of $J_{\ba, \bb}$ defined in (\ref{Jab}) holds true for all indices $\ba, \bb$ for any value of $r,s$. We now wish to prove by induction that the decomposition property will then also hold for all indices $\ba, \bb'$ for all $r$, and $ s $ replaced by $s+1$. The corresponding $J$-function is given by
\bea
J_{\ba, \bb'} (\bx, \by') 
=
\sum _{{0 < m_1 , \dots , m_r \atop 0 < n_1, \dots, n_s, n_{s+1}}}
\delta \left ( \sum _{i=1}^r m_i - \sum _{j=1}^s n_j - n_{s+1} \right )
\prod _{i=1}^r { x_i^{m_i} \over m_i ^{a_i}} 
\prod _{j=1}^{s+1} { y_j^{n_j} \over n_j ^{b_j}} \, .
\eea
and we use the notation $\bb' =(b_1 , \dots, b_s, b_{s+1})$ and $\by ' = (y_1 , \dots, y_s, y_{s+1})$.

\sm

To solve the Kronecker $\delta$-function constraint now requires more care than in the case $s=1$. One cannot simply use the solution $n_{s+1} = m_1 + \dots m_r - (n_1 + \dots +n_s)$, since then $n_{s+1}$ is not guaranteed to obey the constraint that it must be positive. Therefore, we shall proceed as follows instead. We single out an arbitrary $m$-variable, say $m_1$, and compare the values of $n_{s+1}$ allowed by the $\delta$-function constraint to the value of $m_1$. We have three cases, namely $n_{s+1}= m_1$,
$n_{s+1} < m_1$ or $n_{s+1} > m_1$, and split the sum which defines the function $J$ accordingly into three terms
\bea
J_{\ba, \bb'} (\bx, \by')  = J^0 + J^+ + J^- \, ,
\eea
where the terms $J^0,  J^+ $, and $ J^-$ respectively include the contributions to the sum from $n_{s+1}= m_1$, $n_{s+1} < m_1$, and  $n_{s+1} > m_1$.

\sm

For the contributions to the sum with $n_{s+1} < m_1$, which are regrouped in $J^+$, we parametrise the difference by $m_1 = n_{s+1} + k$ with $k$  ranging over all positive integers. The sum becomes, 
\bea
J^+  =
\sum _{{0 < k, m_2 , \dots , m_r \atop 0 < n_1, \dots, n_s, n_{s+1}}}
\delta \left ( k+ \sum _{i=2}^r m_i - \sum _{j=1}^s n_j  \right )
{ x_1 ^{k+n_{s+1}} \over (k+n_{s+1})^{a_1} } \prod _{i=2}^r { x_i^{m_i} \over m_i ^{a_i}} 
\prod _{j=1}^{s+1} { y_j^{n_j} \over n_j ^{b_j}} \, .
\eea
We re-baptise the variable $k$ as $m_1$, setting $k=m_1$ in the sum. As a result, we have
\bea
J^+ =
\sum _{{0 < m_1 , \dots , m_r \atop 0 < n_1, \dots, n_s, n_{s+1}}}
\delta \left ( \sum _{i=1}^r m_i - \sum _{j=1}^s n_j  \right )
{ x_1 ^{m_1 +n_{s+1}} \over (m_1+n_{s+1})^{a_1} } \prod _{i=2}^r { x_i^{m_i} \over m_i ^{a_i}} 
\prod _{j=1}^{s+1} { y_j^{n_j} \over n_j ^{b_j}} \, .
\eea
Note that the $\delta$-function constraint involves only the first $r+s$ summation variables but does not involve $n_{s+1}$. This observation is crucial to make the inductive proof possible. The summations over $m_1$ and $n_{s+1}$ are coupled only through the denominator factor $(m_1+n_{s+1})^{-a_1}$ but not through the Kronecker $\delta$-function constraint. To complete the inductive argument, we decouple the summations over $m_1$ and $n_{s+1}$ by differentiating in $x_1$. The independent sums are then easily regrouped, and lead to the following differential equation
 \bea
\left ( x_1 { \p \over \p x_1} \right )^{a_1} J^+ =  \Li_{b_{s+1}} (x_1 y_{s+1} ) 
\Li_{0, a_2, \dots , a_r, b_1 , \dots , b_s} (x_1, \dots, x_r, y_1 , \dots, y_s) \, .
\eea
Stuffle identities (\ref{stuffle}) may again be used to decompose this product into a linear combination, with rational coefficients, of ${\rm Li}$-functions of depth $r+s+1$. The above differential equation for $J^+$ may then be integrated iteratively using (\ref{e:DiffMpl}) in terms of linear combinations of ${\rm Li}$-functions of depth $r+s+1$. This concludes the part of the proof for the contributions $J^+$ to $J_{\ba, \bb}(\bx, \by)$. 

\sm

For the contributions to the sum with $n_{s+1} > m_1$, which are  regrouped in $J^-$, we parametrise the difference by $m_1 + k = n_{s+1} $ with $k$  ranging over all positive integers. The arguments now proceed as in the case of $J^+$ with labels of positive and negative powers of  $\zeta$ reversed.

\sm

Finally, for the contributions to the sum with $n_{s+1}=m_1$, which are regrouped in~$J^0$, we have
\bea
J^0
=
\sum _{{0 < m_1 , \dots , m_r \atop 0 < n_1, \dots, n_s, n_{s+1}}}
\delta \left ( \sum _{i=2}^r m_i - \sum _{j=1}^s n_j \right )
{ (x_1 y_{s+1}) ^{m_1} \over m_1 ^{a_1+b_{s+1}}}
\prod _{i=1}^r { x_i^{m_i} \over m_i ^{a_i}} 
\prod _{j=1}^s { y_j^{n_j} \over n_j ^{b_j}}  \, .
\eea
The sum over $m_1$ is now completely decoupled from the other sums, and we have
\bea
J^0
=
\Li _{a_1+b_{s+1}} (x_1 y_{s+1}) \, \Li _{a_2, \dots, a_r, b_1 , \dots , b_s} (x_2, \dots, x_r, y_1 , \dots, y_s)
\eea
Applying the stuffle relations of (\ref{stuffle}) allows us to decompose this product again into a linear combination with rational coefficients of ${\rm Li}$-functions. This completes the proof for all contributions to $J_{\ba, \bb}(\bx, \by)$, and thus of this function itself.

\subsection{Arguments for the $v$-integral in the general case} 

While the proof of the \claim\, given in the preceding section is complete   for star graphs, and graphs derived from it, we need further results to support the \claim\, for general graphs. In this subsection, our aim will not be to provide a complete proof for general graphs, but rather to show that some of the key novel results that are needed for general graphs but were not required for star graphs, can be established in a satisfactory way.

\sm

One formulation may be obtained by considering the $v$-integral of a general multiple polylogarithm, but with arguments given by higher powers of $\zeta$ and $\zeta ^{-1}$ (for star graphs only arguments linear in $\zeta$ and $\zeta ^{-1}$ were required). This will lead to a general $v$-integral of the form
\bea
\cJ^{\gamma} _{ \ba} (\bz)
= \int ^1 _0 dv \, \Li _{\ba} ( z_1 e^{2 \pi i \gamma _1 v} , \dots, z_r e^{2 \pi i \gamma _r v} ) \, ,
\qquad
\eea
where the exponent $\gamma = (\gamma _1, \dots, \gamma _r)$ has integer entries, while $\ba = (a_1, \dots, a_r)$ has positive integer entries, and the arguments are again bounded  by $|z_i| <1$ for all $i=1,\dots, r$. By expanding the  ${\rm Li}$-function in a power series in $z_i$, we may readily perform the integral over $v$ to obtain
\bea
\cJ^{\gamma} _{ \ba} (\bz) = \sum _{0 < \mu_1 < \dots < \mu_r } 
\delta \left ( \sum _{i=1}^r \gamma _i \mu_i \right ) 
\prod _{i=1}^r { z_i ^{\mu_i} \over \mu_i ^{a_i}}  \, ,
\eea
where the summation is over ordered positive integers $\mu_1, \dots, \mu_r$.
The case considered earlier corresponds to having $\gamma _i=\pm 1$, so that  the 
Kronecker $\delta$-function constraint may be solved to provide an iterative proof as we did in the preceding subsection. When $\gamma _i \not =\pm 1$, however, one needs to deal with the issues of divisibility in solving the constraint equation imposed by the Kronecker $\delta$-function.

\sm

We begin by parametrising  the ordering conditions on the indices $\mu_i$ by setting $\mu_0=0$ and  introducing the following  change of variables
\bea
\label{mi}
\mu_i = \mu _{i-1} + m_i  \hskip 0.6in 0 < m_i \hskip 0.5in 1 \leq i \leq r \, .
\eea
The variables $m_i$ are now constrained only to be positive, but no further ordering 
requirement between them is being imposed. Importantly, this change of variables has the effect of decoupling the ordering conditions between different $m_i$. Next, it will be convenient to express $\gamma$ and $\bz$ in this new basis  and we define
\bea
\alpha _i = \sum _{j=i}^r \gamma _j 
\hskip 1in
x_i = \prod _{j=i} ^r z_j \, .
\eea
As a result, the function $\cJ^\gamma _\ba (\bz)$ admits the following sum representation
\bea
\cJ^\gamma _\ba (\bz)
=
\sum _{0 < m_1, \dots, m_r} \delta \left ( \sum _{i=1}^r \alpha _i m_i \right ) 
\prod _{i=1}^r { x_i ^{m_i} \over \mu_i ^{a_i}}  \, .
\eea
To save unnecessary extra notation, the denominator factors have been left in terms of $\mu_i$
and should be viewed as given in terms of the functions of the $m_i$ by (\ref{mi}). Without loss of generality, we shall assume that all $\alpha _i $ are non-vanishing; in the contrary case, the corresponding summation indices do not enter  into the Kronecker $\delta$-function constraint and the sum reduces to a case with lower value of $r$. 

\sm

Next, we partition the set  of indices $\{ 1, 2, \dots , r\}$ into two sets $I_+$ and $I_-$ according to 
whether $\alpha _i$ is positive or negative
\bea
\alpha _i >0 & ~ \Longrightarrow ~ & i \in I_+
\no \\
\alpha _i < 0 & ~ \Longrightarrow  ~ & i \in I_-  \, .
\eea
It will be convenient to work with positive exponents only, so we shall set $\beta _i = - \alpha _i $ whenever $i \in I_-$, so that  $\beta _i >0$. The summation may then be rearranged as follows
\bea
\cJ^\gamma _\ba (\bz)
=
\sum _{0 < m_1, \dots, m_r} \delta \left ( \sum _{i\in I_+ } \alpha _i m_i - \sum _{i \in I_-} \beta _i m_i \right ) 
\prod _{i=1}^r { x_i ^{m_i} \over \mu_i ^{a_i}}  \, .
\eea
We now proceed iteratively on the value of $n$, and assume that the extended array 
$\gamma ' = (\gamma _1, \dots, \gamma _r, \gamma _{r+1})$ produces one additional 
entry  belonging to $I_-$, which corresponds to an additional exponent  $\beta _{i_-}>0$ 
and an additional summation variable $m_{i_-}>0$.

\sm

We now wish to reduce the Kronecker $\delta$-function involving $n+1$ summation variables $m$ to one that involves only $n$ summation variables, imitating the procedure we followed for star graphs. To this end, we single out an index $i_+$ in $I_+$ corresponding to $m_{i_+} \alpha _{i_+}$. Thus, we need to reduce the following constraint on $n+1$ summation variables $m_1 , \dots, m_{n+1}$
\bea
\sum _{i \in I_+ \setminus \{ i_+ \}}  \alpha _i m_i 
- \sum _{i \in I_- } \alpha _i m_i + \alpha _{i_+} m_{i_+} - \beta _{i_-} m_{i_-} =0 \, ,
\eea
to a summation involving only $n$ summation variables. We proceed as in the case of star graphs, 
but take careful account of the divisibility issues. We shall distinguish contributions for which 
$\alpha _{i_+} m_{i_+} = \beta _{i_-} m_{i_-} $ from those for which $\alpha _{i_+} m_{i_+} > \beta _{i_-} m_{i_-} $ and those for which $\alpha _{i_+} m_{i_+} < \beta _{i_-} m_{i_-} $. We shall refer to these 
partial sums respectively as $\cJ^0, \cJ^+$, and $\cJ^-$. 

\sm

We shall concentrate here on the contribution $\cJ^+$, arising from the contributions satisfying 
$\alpha _{i_+} m_{i_+} > \beta _{i_-} m_{i_-} $. The other cases are analogous, following the model of the star graphs. The integers $\alpha _{i_+}$ and $\beta _{i_-}$ are both positive, and have been assumed to be non-zero. We shall denote their greatest common divisor by $\alpha _\star >0$. There now exists a unique solution $m^0 _{i_\pm}$ to the B\'ezout equation
\bea
\alpha _{i_+} m^0 _{i_+} - \beta _{i_-} m^0 _{i_-} = \alpha _\star
\eea
with the properties $0 \leq m_{i_+}^0 < \beta _{i_-} /\alpha _\star$ and 
$0 \leq m_{i_-}^0 < \alpha _{i_+}/\alpha _\star$.
Note that this solution is determined completely by the nature of $\alpha _{i_+}$ and $\beta _{i_-}$ and does not in any way involve the summation variables $m$. Having determined $m^0 _{i_\pm}$, we may now 
recast the constraint as follows
\bea
\sum _{i \in I_+ \setminus \{ i_+ \}} \alpha _i m_i  - \sum _{i \in I_- } \alpha _i m_i +    \alpha _\star k =0
\eea
for an integer $k$ which is positive by the assumption $\alpha _{i_+} m_{i_+} > \beta _{i_-} m_{i_-} $.
As a result, the Kronecker $\delta$-function constraint now involves only $n$ summation variables. The complete parametrisation of the summation variables $m_{i_\pm}$ is given 
in terms of the integer $k > 0$, and an additional integer summation variable $\ell$,  as follows
\bea
m_{i_+} & = & k \, m^0 _{i_+} + \ell \, { \beta _{i_-} \over \alpha _\star}
\no \\
m_{i_-} & = & k \, m^0 _{i_-} + \ell \, { \alpha _{i_+} \over \alpha _\star}
\eea
The variable $\ell $ does not enter into the constraint. In general, the range of $\ell$ will include both positive, zero, and negative integers, and must itself be subdivided into these three sub-ranges. Within each sub-range, the summation over $\ell$ may be performed, after sufficient derivatives in the external variables $x_i$ have been taken to deal with the denominator factors $\mu_i ^{a_i}$, in a manner analogous to how we proceeded for star graphs.  
This concludes our discussion of the arguments needed to prove that the integral over $v$ produces a linear combination, with rational coefficients, of Li-functions.

\subsection{Integral over the variable $u$}
\label{sec:uint}

The second type of integral involves integration over one of
the variables $u_i$, leading to the integrals of the type
\begin{equation}
\cK ^\alpha_{a,  \ba} (\bx) =  \int_0^1 du \, u^a\,  \Li_{a_1,\dots,a_r}(x_1 e^{-2\pi \tau_2 \alpha_1 u},
  \dots ,x_r e^{-2\pi  \tau_2 \alpha_r u})  \,,
\end{equation}
where the  variables $x_i$ are independent of $u$ and bounded by $|x_i|<1$, while $a$, $a_i$ and $\alpha_i$ are positive or zero  integers for all $i=1,\dots, r$. We have introduced the notation $\alpha =(\alpha_1, \dots, \alpha _r)$. Actually, the integrals make perfect sense when $\alpha _i$ is a positive integer, and it will sometimes be convenient for performing intermediate calculations, to let $\alpha _i$ be real.  The multiple polylogarithm arises from stuffle relations in the expansion of products of Green functions on the torus, while the pre-factor $u^a$ arises from expanding the Bernoulli polynomials in $D_{a_i,a_i}(\zeta)$. 

\smallskip

It is sufficient to study the integrals $\cK_{0,\ba}^\alpha $, since cases with $a>0$ may be obtained by differentiating with respect to one the arguments $\alpha _i$ (temporarily allowing it to be a real variable instead of an integer) of the polylogarithms using the differentiation rule of the multiple polylogarithms in~\eqref{e:DiffMpl}. Using the sum representation of the multiple polylogarithms and the integral
\begin{equation}
 \int_0^1 du \, e^{-2\pi
    \tau_2 u\, \alpha}
={1\over2\pi \tau_2}\, 
  {1\over \alpha}
 \left(1-e^{-2\pi \tau_2 \alpha}\right)
\end{equation}
gives
\begin{equation}
\cK_{0,\ba}^\alpha (\bx) =  {1\over 2\pi \tau_2} \sum_{0 < m_1<\dots <m_r} \prod_{i=1}^r
  {x_i^{m_i}\over m_i^{a_i}}\,  {1-e^{-2\pi \tau_2 \sum_{i=1}^r \alpha_i
      m_i}\over \sum_{i=1}^r \alpha_i m_i}\,.
\end{equation}
By differentiation we obtain
\begin{multline}
\label{e:diffJ1}
  \left(\sum_{i=1}^r \alpha_i x_i{d\over dx_i}\right) \cK_{0,\ba}^\alpha (\bx) =
  {1\over2\pi\tau_2}\, \Big(\Li_{a_1,\dots,a_r}(x_1,\dots,x_r)\cr
-    \Li_{a_1,\dots,a_r}(x_1 e^{-2\pi \tau_2\alpha_1},\dots,x_r e^{-2\pi \tau_2\alpha_r})\Big)  \,.
\end{multline}
Again, using the differential relations of multiple
polylogarithms~\eqref{e:DiffMpl} one  concludes that
 $\cK_{0,\ba}^\alpha $ is a combination of multiple polylogarithm of
depth $r$ and weight $\sum_{i=1}^r a_i+1$.
The same argument implies that $\cK_{a,\ba}^\alpha $ is a linear combination of 
multiple polylogarithms of depth $r$ and weight $\sum_{i=1}^r a_i+a$.

%%%%%%%%%%%%%%%%%%%%%%%%%%%%%%%%%%%%%%%%%%%%%
%%%%%%%%%%%%%%%%%%%%%%%%%%%%%%%%%%%%%%%%%%%%%
\section{Fourier coefficients of modular graph functions}
\label{sec:6}
%%%%%%%%%%%%%%%%%%%%%%%%%%%%%%%%%%%%%%%%%%%%%
%%%%%%%%%%%%%%%%%%%%%%%%%%%%%%%%%%%%%%%%%%%%%

Having proved the \claim\,  for the case of star graphs, and graphs derived from star graphs, and presented an outline of the evidence for the validity of the \claim\,  for general graphs, we now proceed to discuss some implications of the \claim, assuming it holds true for general graphs. 

\subsection{Implications for the constant Fourier mode}

Assuming the validity of the conjecture implies the following corollary of the conjecture concerning the constant Fourier mode of the Feynman integral $I_\Gamma (q;1)=I_\Gamma (q)$.\\

\noindent {\bf Proposition}

\sm

{\it  Let $I^0_\Gamma (\tau_2)$ denote the constant Fourier mode of the modular graph function
$I_\Gamma (q)$  associated with a graph $\Gamma = \Gamma _{w,N}$ with $w$  Green functions 
and $N$ vertices 
\begin{equation}
I^0_\Gamma (\tau_2)= \int_{-\frac12}^{\frac12} \, d \tau _1 \, I_\Gamma (q)  \,.
\end{equation}
\begin{enumerate}
\itemsep = 0.08in
\item The constant Fourier mode $I^0_\Gamma (\tau_2)$  has an expansion for large $\tau_2$  consisting  of a  Laurent polynomial  with a term of  highest degree  $(\pi \tau_2)^{w}$ and a term of lowest degree $(\pi \tau_2)^{1-w}$ with exponentially suppressed corrections of order ${\mathcal O}  (e^{-2 \pi \tau_2})$. 
\item The coefficients of the $(\pi \tau _2)^k$ term ( $w \leq k \leq w-1$)   in the Laurent polynomial  
is a  single-valued multiple zeta value of weight $w-k$.
\end{enumerate}
}

\bigskip

The first item is straightforward to establish directly from the definition of the modular graph functions. The second item follows from the Conjecture since the modular graph function $I_\Gamma(q)$  results from evaluating a  linear combination, with rational coefficients, of multiple polylogarithms.
The coefficient are single-valued multiple zetas  as a consequence of the single-valueness with respect to $\zeta$ of the modular graph function $I_\Gamma(q,\zeta)$ in eq.~\eqref{e:IGammaDefzeta}. Since
single-valued multiple polylogarithms evaluated at unit argument yield
single-valued multiple zeta functions, the validity of the \claim\ 
supports the validity of item (2) above. Furthermore, item (1) also
follows from item (2) since there are no single-valued multiple zeta
values of negative weights.  By convention a  single-valued multiple zeta
of weight zero is a rational number.

\sm

We shall now provide examples of graphs, and their associated modular graph functions, for which these Laurent polynomials of the constant Fourier mode  have been calculated, or for which the general structure
of the corresponding Laurent polynomial is understood on general grounds.

\subsection{The Eisenstein series}
\label{sec:aprop}

The simplest case where the implications of the Corollary on the constant Fourier mode may be easily verified is the non-holomorphic Eisenstein series itself. The constant Fourier mode consists of  a Laurent polynomial in $\tau_2$ with just two terms. Throughout, it will be convenient to give the Laurent series in powers of $y=\pi \tau_2$ instead of $\tau_2$, given the structure of the coefficients announced in item (3) of the implications.
For the Eisenstein series, we have
\begin{equation}
  \int_{-\frac12}^{\frac12} d \tau_1 \, E_a(q)  = (-1)^{a-1}
  {B_{2a}\over(2a)!}(4y)^a+{4(2a-3)!\over (a-2)!(a-1)!}\zeta(2a-1) (4y)^{1-a}\,.
\end{equation}
The coefficients of powers of $y$ are rational numbers multiplying odd $\zeta$-values  
which are single-valued according the definition used in~\cite{brownCRAS,Schnetz:2013hqa,Brown:2013gia}. 
 The leading coefficient is a rational number involving the Bernoulli number $B_{2a}$, which may also be described as ``single-valued'' since the Bernoulli polynomials in (\ref{e:polybern}) are single-valued.
This exemplifies the relation between the world-sheet graphs and the single-valued elliptic multiple
polylogarithms in the very simplest cases.

\subsection{The $C_{a,b,c}(q)$ modular graph functions}

Although explicit expressions for the Laurent series of the general two-loop modular graph $C_{a,b,c}$
are not available in closed form, it is not difficult to verify the compatibility of their structure with 
the Corollary given above. First of all, the Laurent series of every specific $C_{a,b,c}$ function evaluated explicitly  in  \cite{D'Hoker:2015foa} is of the form predicted by the Corollary, with the highest power being $y^w$ and the lowest $y^{1-w}$. The simplest case is $C_{1,1,1}=E_3 +\zeta (3)$, which is manifestly compatible, in view of the Fourier mode expansion of the Eisenstein series given in the preceding section. 

\sm

The cases of weight 4 and 5 worked out in \cite{D'Hoker:2015foa} give the following Laurent expansions
\bea
\int _{- \half} ^\half \!\! d \tau _1 C_{2,1,1} (q)   \! \!\!\! & = &  \!\!\! \! 
\frac{2   y^4 }{14175}  +  \frac{ \zeta(3) y}{45}  + \frac{5  \zeta(5)}{12 y}  -\frac{ \zeta(3)^2}{4 y^2} +  \frac{9 \zeta(7)}{16  y^3}+{\mathcal O}\left(e^{-2y}\right) ,
\\
\int _{- \half} ^\half   \!\! d \tau _1  C_{3,1,1} (q)
 \! \!\!\! & = & \! \!\!\!
 \frac{2  y^5}{155925}  +\frac{2   \zeta (3) y^2 }{945} 
   -\frac{\zeta (5)}{180}    +\frac{7 \zeta (7)}{16 y^2}
 -\frac{\zeta (3)  \zeta (5)}{2 y^3} +\frac{43  \zeta (9)}{64 y^4}+{\mathcal O}\left(e^{-2y}\right) ,
\no
\eea
and these expressions are again compatible with the structure
predicted by the Corollary since the coefficients only involve
polynomials in odd $\zeta$ values, which are single-valued multiple zetas.

\sm

This structure of the Laurent series is easily seen to follow from the structure of nested differential equations satisfied by the~$C_{a,b,c}(q)$ functions. Indeed, every Laplace-eigenvalue equation of \cite{D'Hoker:2015foa} for linear combinations (with rational coefficients) $\mC_{w,s,\mp}$ of $C_{a,b,c}$-functions of weight $w=a+b+c$ and eigenvalue $s=1,2, \dots, w-2$, has an inhomogeneous part consisting of a term linear in the Eisenstein series $E_w$ multiplied by a rational number plus terms of the form $E_{w_1} E_{w_2}$ with $w_1, w_2 \geq 2$ and $w=w_1+w_2$, and multiplied by rational coefficients. Therefore, the inhomogeneous part  involves a Laurent polynomial with powers  $y^k$ ranging from $1-w \leq k \leq w$, the coefficient of the term $y^w$ being rational, and the remaining coefficients being odd $\zeta$-values, which by definition are proportional to single-valued zeta values with rational coefficients. We may represent this by
\bea
\left ( y^2 { \p^2 \over \p y^2} -s(s-1) \right ) \mC_{w,s,\mp} \,  \in \,
 \IQ E_w + \sum _{{w_1, w_2 \geq 2 \atop  w_1+w_2=w}} \IQ E_{w_1} E_{w_2} \, .
\eea
The operator on the left hand side of the equation has a kernel
consisting of the  Laurent terms proportional to $y^s$ and $y^{1-s}$.
The Laplace eigenvalue equations therefore imply that almost all the
coefficients  in the Laurent polynomial for $\mC_{w;s;\mp}$, except
for the highest coefficient of $y^w$, are odd zeta values or products
thereof, as indeed predicted by the Corollary.  

\subsection{Examples of higher cases}

Some Laurent expansions of modular graph functions that do not belong
to the family of graphs $C_{a,b,c}$ were also evaluated in
\cite{D'Hoker:2015foa}. Here, we give just two examples of Laurent
series at weight 5, with $y=\pi \tau_2$
\bea
\int _{- \half} ^\half \!\! d \tau _1  D_{3,1,1} (q) \!\!\! \!  & = & \!\!\! \! 
\frac{2y^5}{22275}
+ \frac{y^2 \zeta(3)}{45}
+ \frac{11  \zeta(5)}{60}  
+ \frac{105  \zeta(7)}{32 y^2}
- \frac{3  \zeta(3) \zeta(5)}{2 y^3} 
+ \frac{81 \zeta(9)}{64 y^4}  + {\mathcal O}\left(e^{-2y}\right) ,
\no \\ 
\int _{- \half} ^\half \!\! d \tau _1 D_{2,2,1}  (q)
\!\!\! \!  & = & \!\!\! \! 
\frac{8 y^5}{467775}
+\frac{4   \zeta (3) y^2}{945}
+\frac{13 \zeta (5)  }{45} 
+ \frac{7 \zeta(7) }{8  y^2} 
- \frac{\zeta(3)\zeta(5) }{y^3}  
+ \frac{9  \zeta(9) }{8 y^4}+{\mathcal O}\left(e^{-2y}\right) . \nn\\
 \eea
In each case, only odd zeta values, or products thereof, occur all of which are single-valued multi-zeta values.  The preceding  expressions  do not involve non-trivial single-valued multiple zetas  (i.e. ones which are not reducible to polynomials in odd zeta values) since these only arise at weights $\ge 11$. 

\sm

Explicit computations of Laurent series at higher weight have been performed recently by  Zerbini~\cite{Zerbini}.   These strikingly confirm the conjectured structure of the coefficients of the Laurent polynomials.  For example,  Zerbini's evaluation of  the constant Fourier mode of $D_{1,1,5}(q)$ (which is a special case of the class of  star graphs) is given by
\begin{multline}
 { 1 \over 4^7} \int_{-\frac12}^{\frac12} d \tau_1 D_{1,1,5}(q)
 =
 {62 \, y^7 \over10945935} 
 +{\zeta_{sv}(3)\over243} y ^4
 +{119\over648}  \zeta_{sv}(5)y^2 
+{11\over54}\zeta_{sv}(3)^2  y 
\cr
+{21\over32}\zeta_{sv}(7) 
+{23\over} {\zeta_{sv}(3)\zeta_{sv}(5)\over 6y }
+{7115\zeta_{sv}(9)-900\zeta_{sv}(3)^3\over576 y^2} 
\cr
+{1245\zeta_{sv}(3)\zeta_{sv}(7)-150\zeta_{sv}(5)^2\over64 y^3}\cr
+{288\zeta_{sv}(3,5,3)-4080\zeta_{sv}(5)\zeta_{sv}(3)^2-9573\zeta_{sv}(11)\over256 y^4}\cr
+{2475\zeta_{sv}(5)\zeta_{sv}(7)+1125\zeta_{sv}(9)\zeta_{sv}(3)\over128 y^5}
-{1575\over64} {\zeta_{sv}(13)\over y^6}
 +{\mathcal O} (e^{-2 y})\,.  
\end{multline}
The fact that the coefficient of $y^{-4}$ is a linear combination of
depth 3 single-valued multiple zeta values is a non-trivial result. Similarly, the coefficients of $y^{-3}$ and $y^{-5}$  in $D_{1,1,6}$  are linear combinations with rational coefficients of  single-valued multiple zeta of weight 11 and 13 respectively~\cite{Zerbini}.

\smallskip

Another important aspect of the structure  described in this paper is that the modular graph  
functions that arise at special values of the single-valued elliptic multiple polylogarithms can be expressed as integrals of sums 
of single-valued multiple polylogarithms.  This means that their functional form can be 
determined by using the algebraic properties of multiple polylogarithms that were described 
in section~\ref{sec:overview}. 
Although this is an interesting comment on the structure of these functions, it is quite 
complicated to  evaluate the functions in this manner, even for the simplest nontrivial 
function which is $C_{1,1,1}(q)$, which is evaluated explicitly in  appendix \ref{sec:C111}.  
The result has previously been obtained by Zagier (unpublished) 
and in \cite{D'Hoker:2015foa} by other methods.

%%%%%%%%%%%%%%%%%%%%%%%%%%%%%%%%%%%%%%%%%%%%%
%%%%%%%%%%%%%%%%%%%%%%%%%%%%%%%%%%%%%%%%%%%%%
\section{Summary and thoughts on the basis of modular graph functions}
\label{sec:highern}
%%%%%%%%%%%%%%%%%%%%%%%%%%%%%%%%%%%%%%%%%%%%%
%%%%%%%%%%%%%%%%%%%%%%%%%%%%%%%%%%%%%%%%%%%%%

The preceding discussion concerns properties of modular graph functions that are  
associated with  Feynman diagrams for a free conformal scalar field on a torus.  A general graph, 
denoted $\Gamma = \Gamma_{w,N}$,  has $w$ Green functions joining $N$ vertices at positions 
$\zeta_i$ ($i=1,\dots,N$) that are integrated over the torus.   Each Green function is itself a 
single-valued elliptic polylogarithm, $D_{1,1}(q,\zeta_i/\zeta_j)$.  This procedure defines a modular 
function, $I_\Gamma (q)$,  associated with the graph as in \eqref{e:IGammaDef}.   

\smallskip

We also considered the graph with $N+1$ vertices that is obtained from $\Gamma = \Gamma_{w,N}$  
by integrating over only $N-1$ vertices and separating the Green functions joined to 
the $N$-th vertex into two groups groups  that end at the points $\zeta_N$ and $\zeta_{N+1}$.  
This is associated with the function   $I_\Gamma (q;\zeta)$  \eqref{e:IGammaDefzeta}, 
which is a single-valued function of the argument $\zeta=\zeta_{N+1}/\zeta_N$.  
This  is a particular example of a single-valued elliptic multiple polylogarithm.   
The value of this function at the point $\zeta=1$ is a modular graph function, 
$I_\Gamma (q;1)= I_\Gamma (q)$.     

\smallskip

 As in the case of multiple zeta values the question of determining a basis of modular graph functions 
 seems daunting.  In considering this question we should bear in mind that  the context that motivated 
 our analysis was limited.  The  general Feynman diagrams with $N$ vertices and $w$ Green functions 
 that  were considered  in this paper and in
 \cite{D'Hoker:2015zfa,D'Hoker:2015foa,Basu:2015ayg} generalise those
 obtained by  expanding the four-graviton superstring scattering
 amplitude, which only generates the diagrams with  $N\le 4$.   This
 does not provide a complete set of basis functions.  In fact, the
 relations between the functions that were  proved or conjectured in
 \cite{D'Hoker:2015foa} explicitly involve relations between functions
 with at most four vertices and those with more than four vertices.
 
 \smallskip 
  
The preceding comments parallel known properties of  the  expansion
of  the tree-level $N$-graviton amplitude, which generates a subset of
single-valued multi-zeta values at any weight. For example for $N=4$ the expansion coefficients
  are  polynomial in ordinary odd Riemann zeta values.  In order to access the complete basis of multiple zeta values it is necessary to expand the tree amplitudes for general $N$ \cite{Stieberger:2013wea,SCHLotterer:2012ny}.
By analogy,  we would expect  that  a more complete understanding of
the basis for the space of modular graph functions  requires analysis
of the low energy expansion of $N$-graviton genus-one amplitudes, which
is presently rather limited. It is known \cite{Green:2013bza} that for
$N=5$  new kinds of modular graph functions appear for weights $w>4$.  These are graphs in which $2N-8$ Green functions contain a holomorphic or anti-holomorphic  numerator momentum factor, which have the form 
\bea
D_{1,0}(q;\zeta) = -\zeta \,\partial_\zeta D_{1,1}(q;\zeta)\,,\qquad D_{0,1}(q;\zeta) =- \bar \zeta\,\partial_{\bar\zeta} D_{1,1}(q;\zeta)\,.
\label{e:Spinormod}
\eea
These functions transform as modular forms of non-zero modular weight as seen from \eqref{Dabtrans}.
Such graphs  are integrals of products of 
$D_{a,b}(q;\zeta)$ with $a,b=0,1$, with the constraint that  the  integrand has zero net modular weight.  For example the integrands of the modular graph functions of this type in the $N=5$ case have one  $D_{0,1}$ factor and one $D_{1,0}$ factor.

\smallskip
  
Since the relationships between different modular graph functions are difficult to determine with present techniques, it is not clear how many independent modular weight functions there are at any weight.

 \smallskip
The fact that the coefficients of the Laurent polynomials of the constant Fourier mode of modular graph  functions are  single-valued multiple zeta values suggests that they may be related to coefficients in the low energy expansion of tree-level amplitudes. After all, in the limit $\tau_2\to \infty$ the genus-one $N$-graviton  amplitude degenerates to a $N+2$-particle tree amplitude at special values of the momenta.   It would therefore be interesting to connect these coefficients to the single-valued multiple zetas that appear in the low energy  expansion of tree-level closed string amplitudes~\cite{Stieberger:2013wea,Stieberger:2014hba}.

\smallskip

 A related issue is a possible connection between modular graph functions and holomorphic elliptic multiple polylogarithms of \cite{brownlevin}. The latter arise in the low energy expansion of open string theory, as discussed in \cite{Broedel:2014vla}. Such a connection would be analogous to the relation of  ordinary multiple polylogarithms to single-valued multiple polylogarithms.   Within string theory this corresponds to the KLT relation that builds the tree level closed string amplitudes  from the open string tree amplitudes.   There are strong hints that the integrands of the open string one-loop amplitude and the closed string genus-one amplitude considered in this paper are related in an analogous manner.  Determining the details of  such a genus-one ``KLT'' relation would be very enlightening.

%%%%%%%%%%%%%%%%%%%%%%%%%%%%%%%%%%%%%%%%%%%%%
%%%%%%%%%%%%%%%%%%%%%%%%%%%%%%%%%%%%%%%%%%%%%
\section*{Acknowledgments}
%%%%%%%%%%%%%%%%%%%%%%%%%%%%%%%%%%%%%%%%%%%%%
%%%%%%%%%%%%%%%%%%%%%%%%%%%%%%%%%%%%%%%%%%%%%

We would like thank Erik Panzer and Don Zagier for many useful discussions and
exchanges. We are very grateful to Federico Zerbini for providing us
with his evaluation of the Laurent polynomials of various graphs prior to publication.  
We would also like to thank both referees of this paper for their detailed and very helpful comments.

\sm

The research of ED was supported in part by  grants from the National Science Foundation (NSF) PHY-13-13986 and PHY-16-19926. The research of PV has received funding the ANR
grant reference QST 12 BS05 003 01, and the CNRS grants PICS number
6430. The research of OG has received funding by
European Research Council (Programme Ideas ERC-2012-AdG 320769
AdS-CFT-solvable) and by the ANR grant StrongInt (BLANC-SIMI-4-2011).
PV is partially supported by   a fellowship funded by the French Government at Churchill College, Cambridge.
MBG and PV acknowledge funding from the European Research Council under the European 
Community's Seventh Framework Programme (FP7/2007-2013) $\/$ ERC grant agreement no. [247252].

\appendix

%%%%%%%%%%%%%%%%%%%%%%%%%%%%%%%%%%%%%%%%%%%%%
%%%%%%%%%%%%%%%%%%%%%%%%%%%%%%%%%%%%%%%%%%%%%
\section{Polylogarithms and Eichler integrals}
\label{sec:covar-deriv}
%%%%%%%%%%%%%%%%%%%%%%%%%%%%%%%%%%%%%%%%%%%%%
%%%%%%%%%%%%%%%%%%%%%%%%%%%%%%%%%%%%%%%%%%%%%

We will  here demonstrate an interesting connection between elliptic polylogarithms and integrals of holomorphic Eisenstein series.  Part of this discussion is contained in~\cite[\S7]{ZagierGangl}. 
\smallskip

To begin with we will introduce a set of derivatives that shows how the
$D_{a,b}(q;\zeta)$ functions of  section~\ref{sec:ellpoly} can be expressed as derivatives of the
polylogarithms and holomorphic Eisenstein series.   
Consider  the differential operator
\begin{equation}
  \label{eq:DD}
  \mathbb D_a = (-1)^{a-1} \sum_{k=0}^{a-1} \,
    \left(2a-2-k\atop a-1\right)\, {(-\log|x|^2)^k\over k!}\, \left(d\over
      d\log x\right)^k\,.
\end{equation} 
Making use of  the differentiation relation  \eqref{e:diffpoly}
one can express $D_{a,a}(x)$ as 
\begin{equation}\label{e:DaaDer}
  D_{a,a}(x)= 2\Ree\left(\mathbb D_a\Li_{2a-1}(x) \right) \,.
\end{equation}
The action of $  \mathbb D_a$ can be promoted  to
the action of a covariant derivative  
\begin{equation}
\label{covdef}
\mathcal  D_a = (-1)^{a-1} \sum_{k=0}^{a-1} \left(2a-2-k\atop a-1\right)\,
  {(4\pi \tau_2 )^k\over k!}  \,\left(d\over d\log q\right)^k\,,
\end{equation}
which maps weight $\ell$ modular forms to weight $k+\ell$ modular forms.
This differential operator can be expressed as the product of covariant
derivatives~\cite[\S7]{ZagierGangl}
\begin{equation}
    \mathcal  D_a=(4\pi\tau_2)^{m-1}\,\left(\partial_{-2}\circ \partial_{-4}\circ \cdots \circ \partial_{2-2a}\right)\,,
\end{equation}
where 
$\partial_k={1\over 2i\pi} \partial_\tau - {k\over 4\pi \tau_2}$.
Equipped with $\mathcal D_a$  one can rewrite~\eqref{e:DaaDer} in the form 
\begin{equation}
\label{daadef}
  D_{a,a}(q;\zeta)
=  2\Ree \mathcal D_a\left[\widetilde G_{2a-1}(q;\zeta)\right] \,.
\end{equation}
where
\begin{equation}
  \label{e:Eicha}
  \widetilde G_{2a-1}(q;\zeta )= \sum_{n\geq0} \left(
  \Li_{2a-1}(q^n\zeta)+ 
  \Li_{2a-1}(q^{n+1}/\zeta) \right)-{
(\log q)^{2a-1}\over (2a)!} B_{2a}(u)\,.
\end{equation}
Setting $\zeta=1$ and using $B_{2a}(1)=-2a\,\zeta(1-2a)$ leads to an expression for the Eisenstein series  as
\begin{equation}\label{e:EdG}
  E_a(q)= {1\over (-4\pi \tau_2)^{a-1}} 2\Ree\cD_a\left[\widetilde G_{2a-1}(q;1)\right]  \,,
\end{equation}
where the function $\widetilde G_{2a-1}(q;1)$ is given by
\begin{equation}
  \label{e:Gmt}
\widetilde G_{2a-1}(q)= \zeta(1-2a) \, {(\log q)^{2a-1}\over
  (2a-1)!}+ \zeta(2a-1) + 2\sum_{n=1}^\infty
\Li_{2a-1}(q^n)\,.
\end{equation}
In fact, $\widetilde G_{2a-1}(q)$ is  an Eichler integral\footnote{ There is a polynomial ambiguity
  $\sum_{i=0}^{2a-2} a_i \, (\log q)^i$ where $a_i$ are
  constants. This polynomial is the period polynomial of the modular
  form as discussed in~\cite{ZagierPeriod}. } that satisfies 
\begin{equation}
  \left({d\over d\log q}\right)^{2a-1} \widetilde G_{ 2a-1}(q)= 2 G_{2a}(q)\,.
  \label{e:diffga}
\end{equation}
The quantity $G_{2a}(q)$ weight $2a$ holomorphic Eisenstein series for $SL(2,\ZZ)$, defined by 
\begin{equation}
  \label{e:Gm}
  G_{2a}(q)={(2a-1)!\over 2 (2i\pi)^{2a}}\, \sum_{(m,n)\neq(0,0)}
  {1\over (m\tau+n)^{2a}}=\frac12 \zeta(1-2a)+ \sum_{n=1}^\infty \,
  n^{2a-1} {q^n\over 1-q^n}
 \,.
\end{equation}
Since $\Li_0(x)={x\over 1-x}$ the $q$-expansion is given by a weight 0
polylogarithm as implied by the relation~\eqref{e:diffga}.  Combining \eqref{daadef} with \eqref{e:diffga} leads to a relation between the non-holomorphic Eisenstein series  $E_a(q)$ and the holomorphic Eisenstein series, $G_{2a}(q)$.

At the end of the next appendix we will use a similar argument to obtain relations between both  the modular functions $E_3(q)$ and $C_{1,1,1}(q)$ and the holomorphic Eisenstein series $G_6(q)$.

%%%%%%%%%%%%%%%%%%%%%%%%%%%%%%%%%%%%%%%%%%%%%
%%%%%%%%%%%%%%%%%%%%%%%%%%%%%%%%%%%%%%%%%%%%%
\section{The $C_{1,1,1}(q)$ modular graph function}
\label{sec:C111}
%%%%%%%%%%%%%%%%%%%%%%%%%%%%%%%%%%%%%%%%%%%%%
%%%%%%%%%%%%%%%%%%%%%%%%%%%%%%%%%%%%%%%%%%%%%

In this section we illustrate how the connection between  modular graph functions  and single-valued multiple polylogarithms can be used to evaluate their functional form.   We will here consider the first nontrivial example, which is the function $C_{1,1,1}(q)$ that is defined by 
\begin{equation}
  C_{1,1,1}(q)=\int_\Sigma  {d^2\log\zeta\over 4\pi^2 \tau_2} \, D_{1,1}(q;\zeta)^3 \,,
\end{equation}
with $D_{1,1}(q;\zeta)$ defined in~\eqref{e:D11Li1} and $\zeta=q^u e^{2i\pi v}$ with $u,v\in[0,1]$.

We now expand the integrand using the stuffle relations of polylogarithms
obtained using the series representation
\begin{multline}\label{e:Liexpand}
\prod_{i=1}^3\Li_{a_i}(x_i)= \Li_{a_1+a_2+a_3}(x_1x_2x_3)+ \sum_{\sigma\in\mathfrak
  S_3}\Li_{a_{\sigma(1)},a_{\sigma(2)},a_{\sigma(3)}}(x_{\sigma(1)},x_{\sigma(2)},x_{\sigma(3)})\cr
+\sum_{i=1}^3 \delta_{\{i,jk\}=\{1,2,3\}} \,\Big(\Li_{a_i,a_j+a_k}(x_i,x_jx_k)+\Li_{a_j+a_k,a_i}(x_jx_k,x_i)\Big)\,,
\end{multline}
where $x_i$ are the arguments of the
polylogarithms that are of the form in the various terms of~\eqref{e:D11Li1}.  The symbol $\mathfrak S_3$  denotes the set of permutations of three elements,
and $\delta_{\{i,jk\}=\{1,2,3\}}$ enforces the
constraint that $(i,j,k)$ is a permutation of $(1,2,3)$.
We apply this identity with $a_i=1,2$ 

\smallskip

As an illustration we evaluate the contribution from the term in the
integrand with only the factors of $\textrm{Li}_2$.  This has the form
\begin{eqnarray}
I_{2,2,2}& =&  {\tau_2^3\over\pi^3}\int_\Sigma  {d^2\log \zeta\over 4 \pi^2 \tau_2}  \left(\Li_2(e^{2i\pi u})
    +\Li_2(e^{-2i\pi u}) \right)^3 \,\cr
&=& {6\tau_2^3\over\pi^3} \Ree\int_0^1 du \, \Li_2(e^{2i\pi u})^2   \Li_2(e^{-2i\pi u}) \,.
\end{eqnarray}
Substituting  the expansion~\eqref{e:Liexpand}
\begin{multline}
\Li_2(e^{2i\pi u})^2    \Li_2(e^{-2i\pi u})= \Li_6(e^{2i\pi u})
+2\Li_{2,2,2}(e^{2i\pi u}, e^{2i\pi u},e^{-2i\pi u})\cr
+2\Li_{2,2,2}(e^{2i\pi u}, e^{-2i\pi u},e^{2i\pi u})
+2\Li_{2,2,2}(e^{-2i\pi u}, e^{2i\pi u},e^{2i\pi u})
\cr
+2\Li_{2,4}(e^{2i\pi u},1)+2\Li_{4,2}(1,e^{2i\pi u})
+\Li_{2,4}(e^{-2i\pi u},e^{4i\pi u})+\Li_{4,2}(e^{4i\pi u},e^{-2i\pi u})\,,
\end{multline}
and using the series expression for the multiple polylogarithms we find
that the only non vanishing contributions are
\begin{eqnarray}
  I_{2,2,2}&=&{6\tau_2^3\over\pi^3}\Ree\int_0^1 du \,  \left(   2\Li_{2,2,2}(e^{2i\pi u}, e^{2i\pi
  u},e^{-2i\pi u})+ \Li_{4,2}(e^{4i\pi u},e^{-2i\pi u})\right)
  \cr
&=&{6\tau_2^3\over\pi^3} \,\Ree\left(2 \sum_{0<m_1<m_2} {1\over m_1^2m_2^2(m_1+m_2)^2}+{1\over4} \Li_6(1) \right)\,.
\end{eqnarray}
The expression in parenthesis is real and is a special value of the
multiple sum
\begin{equation}\label{e:L3def}
  L_{a,b;c}(x,y) = \sum_{m_1,m_2\geq1} {x^{m_1} y^{m_2}\over m_1^a
    m_2^b (m_1+m_2)^c}\,.
\end{equation}
since
\begin{multline}
  L_{a,b;c}(x,y)={1\over2^c} \sum_{0<m} {(xy)^m \over m^{a+b+c}}\cr
+ \sum_{0<m_1<m_2} {x^{m_1} y^{m_2}\over m_1^a
    m_2^b (m_1+m_2)^c}+ \sum_{0<m_2<m_1} {x^{m_1} y^{m_2}\over m_1^a
    m_2^b (m_1+m_2)^c}\,. 
\end{multline}

The multiple sums $L_{a,b;c}(x,y)$ may be reduced to  a linear
combination of multiple polylogarithms as shown in
appendix~\ref{sec:polyred}, with the result 
\begin{equation}
  L_{2,2;2}(x,y)=\Li_{2,4}({\frac {y}{x}},x)+\Li_{2,4}({\frac {x}{y}},y)+2\,
\Li_{1,5}({\frac {y}{x}},x)+2\,\Li_{1,5}({\frac {x}{y}},y)\,,
\end{equation}
and the value at $x=y=1$ is easily obtained using {\tt Hyperint}
routines of~\cite{Panzer:2014caa}
\begin{equation}
 L_{2,2;2}(1,1)=2\Li_{2,4}(1,1)+4\Li_{1,5}(1,1)={\zeta(6)\over3}\,. 
\end{equation}
Therefore 
\begin{eqnarray}
  I_{2,2,2}= 2 \zeta(6)\, {\tau_2^3\over\pi^3}\,.  
\end{eqnarray}
The remaining integrals are performed similarly.
Collecting everything we have
\begin{equation}
  C_{1,1,1}(q)= {2\zeta(6)\over\pi^3}\tau_2^3+\varphi_0(q)+ {\varphi_1(q)\over\tau_2}+{\varphi_2(q)\over\tau_2^2}  \,,
\end{equation}
where
\begin{equation}
  \varphi_0=\zeta(3)+ 4\sum_{n\geq1} \Ree\Li_3(q^n)+12\sum_{n_1,n_2,n_3\geq1}
            \Ree   L_{1,1;1}(q^{n_1+n_3+1}, q^{n_2+n_3+1})\,,
\end{equation}
and
\begin{multline}
  \varphi_1(q)= -{3\over\pi}\sum_{n_1,n_2\geq0} \Ree(
                  \Li_4(q^{n_1}\bar q^{n_2})+\Li_4(q^{n_1+1}\bar
                  q^{n_2+1}))  \cr
+{3\over\pi}\, \sum_{n_1,n_2,n_3>0} \Ree\left(L_{1,1;2}(q^{n_1}\bar
  q^{n_3},q^{n_2}\bar q^{n_3})-L_{1,1;2}(q^{n_1+1}\bar
  q^{n_3+1},q^{n_2+1}\bar q^{n_3+1})\right)\cr
+{6\over\pi} \sum_{n_1,n_2,n_3\geq0} \Ree\left(L_{2,1;1}(q^{n_1}\bar
  q^{n_3},\bar q^{n_2+n_3+1})-L_{2,1;1}(q^{n_1+1}\bar q^{n_3+1},\bar q^{n_2+n_3+1})\right)\,,
\end{multline}
and finally
\begin{equation}
  \varphi_2(q)=-{3\over4\pi^2}\zeta(5)+{3\over2\pi^2}\sum_{n\geq0} \Ree\Li_5(q^n)\,.
\end{equation}

We note that the Eisenstein series  $E_3(q)$ has an expansion that follows from~\eqref{e:EdG} of the form
\begin{multline}
  E_3(q)= {2\zeta(6)\over\pi^3} \tau_2^3+ {3\zeta(5)\over4\pi^2\tau_2^2}+ \sum_{n\geq1}
  \Big(2n^2 \Ree \Li_3(q^n)+{3n\over \pi\tau_2}\Ree
   \Li_4(q^n)\cr
+{3\over2\pi^2\tau_2^2}\Ree \Li_5(q^n)\Big)  \,.
\end{multline}
Putting everything together we have
\be
 C_{1,1,1}(q)-E_3(q)-\zeta(3)=\Ree\left(2 \phi_0(q)  +{3\over\pi} { \phi_1(q)\over\tau_2}\right)\,,
 \label{e:c11res}
\ee
where
\begin{equation}
\phi_0(q)= 6\sum_{n_1,n_2,n_3\geq0} L_{1,1;1}(q^{n_1+n_3+1},
q^{n_2+n_3+1})+\sum_{n\geq1} (1-n^2) \, \Li_3(q^n)\,.
\end{equation}
and 
\begin{multline}
 \phi_1(q)= -\sum_{n\geq1} n \Li_4(q^n) + \sum_{n_1,n_2\geq0}    \left(\Li_4(q^{n_1}\bar q^{n_2})+
    \Li_4(q^{n_1+1}\bar q^{n_2+1})\right)\cr
-\sum_{n_1,n_2,n_3\geq0} \left(L_{1,1;2}(q^{n_1}\bar
  q^{n_3},q^{n_2}\bar q^{n_3})-L_{1,1;2}(q^{n_1+1}\bar
  q^{n_3+1},q^{n_2+1}\bar q^{n_3+1})\right)\cr
-2 \sum_{n_1,n_2,n_3\geq0} \left(L_{2,1;1}(q^{n_1}\bar
  q^{n_3},\bar q^{n_2+n_3+1})-L_{2,1;1}(q^{n_1+1}\bar
  q^{n_3+1},\bar q^{n_2+n_3+1})\right)\,.
\end{multline}
It is striking that these identities are not reducible to standard
polylogarithm identities and that they mix the $q$ and $\bar q$
expansion as is seen in the expression for $\phi_1(q)$.

Using the expressions for $L_{a,b;c}(x,y)$ derived in appendix~\ref{sec:polyred} 
\begin{eqnarray}
L_{1,1;1}(x,y)&=&\Li_{1,2}(y/x,x)+\Li_{1,2}(x/y,y)\cr
  L_{1,1;2}(x,y)&=& \Li_{1,3}(y/x,x)+\Li_{1,3}(x/y,y)  \cr
L_{2,1;1}(x,y)&=&\Li_{2,2}(x/y,y)+\Li_{1,3}(y/x,x)+\Li_{1,3}(x/y,y)\,.
\end{eqnarray}
one can easily check that $\phi_0(q)=0=\phi_1(q)$ to an arbitrary
order in the $q$ expansion using {\tt HyperInt}~\cite{Panzer:2014caa}.
It then follows from \eqref{e:c11res} that 
\bea
C_{1,1,1}(q)=E_3(q)+\zeta(3)\,.
\eea

 \subsection{Remark on  Eichler integrals}\label{sec:eich}

Introducing the covariant derivative 
\begin{equation}
\mathcal  D_2 = -2- 4\pi \tau_2 \,{d\over d\log q}\,,
\end{equation}
we find that
\begin{equation}
  E_3(q)=-{1\over(-4\pi \tau_2)^2} \, 2\Ree \mathcal D_2 \widetilde G_5(q)  
\end{equation}
and
\begin{equation}
  C_{1,1,1}(q)=- {1\over(-4\pi \tau_2)^2} 2\Ree \mathcal D_2\widehat G_5(q)\,,
\end{equation}
where
\begin{equation}
\widetilde G_{5}(q)= \zeta(-5) \, {(\log q)^{6}\over
  5!}+ \zeta(5) + 2\sum_{n=1}^\infty
\Li_{5}(q^n)\,
\end{equation}
and
\begin{equation}
  \widehat G_5(q)= \widetilde G_5(q) +\frac12\pi^3\zeta(3)  (\log q)^2  \,.
\end{equation}
Both of these function satisfy
\begin{equation}
\left(d\over d \log q\right)^5 \widetilde
G_5(q)=\left(d\over d \log q\right)^5 \widehat
G_5(q)= 2G_6(q)\,.
\end{equation}
where $G_6(q)$ is  the holomorphic Eisenstein series
\begin{equation}
  G_6(q)={60\over(2i\pi)^6}\, \sum_{(m,n)\neq(0,0)} {1\over (m\tau+n)^6}  \,.
\end{equation}
This means that $E_3(q)$ and $C_{1,1,1}(q)$ are related to two Eichler
integrals of the  holomorphic weight 6 Eisenstein series for
$SL(2,\ZZ)$. The difference between $\widetilde G_5(q)$ and $\widehat
G_5(q)$ is the polynomial ambiguity related to the period polynomial
arising when integrating the holomorphic Eisenstein series. 
We refer back  to the appendix~\ref{sec:covar-deriv}  for a review of this construction.

%%%%%%%%%%%%%%%%%%%%%%%%%%%%%%%%%%%%%%%%%%%%%
%%%%%%%%%%%%%%%%%%%%%%%%%%%%%%%%%%%%%%%%%%%%%
\section{Reduction of multiple sums to multiple polylogarithms}
\label{sec:polyred}
%%%%%%%%%%%%%%%%%%%%%%%%%%%%%%%%%%%%%%%%%%%%%
%%%%%%%%%%%%%%%%%%%%%%%%%%%%%%%%%%%%%%%%%%%%%

In this appendix we wil reduce various multiple constrained sums to
multiple polylogarithms, making use of  the partial
fractions identity given in~\cite{ZagierDm}  
\begin{equation}
  \label{e:Id3}
  {1\over m^a n^b}= \sum_{r=b}^{a+b-1} {{r-1\choose
        b-1}\over (m+n)^r m^{a+b-r}}+\sum_{r=a}^{a+b-1}
{{r-1\choose
        a-1}\over (m+n)^r n^{a+b-r}} \,.
\end{equation}
These reductions have  been checked with the program {\tt
    HyperInt}~\cite{Panzer:2014caa}.\footnote{ We thank Erik Panzer for help in performing theses checks.}

\smallskip

We will now reduce the multiple constrained sum $L_{a,b;c}(x,y)$ that arose in the evaluation of
$C_{1,1,1}(q)$ to a sum of  multiple polylogarithms, which is given by
\begin{equation}
   L_{a,b;c}(x,y)= \sum_{m_1,m_2\geq1} {x^{m_1}y^{m_2}\over m_1^a
     m_2^b (m_1+m_2)^c}
\end{equation}
can be expressed, when  $a,b>0$, in the form
\begin{equation}\label{e:LabcLi}
  L_{a,b;c}(x,y)= \sum_{r+s=a+b\atop r,s>0} \,\left( {r-1\choose a-1}
    \, \Li_{s,c+r}(y/x,x)+   {r-1\choose b-1}
    \, \Li_{s,c+r}(x/y,y) \right)\,.
\end{equation}
Clearly $L_{a,b;c}(x,y)=L_{b,a;c}(y,x)$ and 
\begin{eqnarray}
  L_{0,b;c}(x,y)&=&\Li_{b,c}(y/x,x)\\  
  L_{a,b;0}(x,y)&=&\Li_a(x)\Li_b(y)\,.
\end{eqnarray}
One can integrate the differential equation 
\begin{equation}
  \left(x{d\over dx}+y{d\over dy}\right)^c L_{a,b;c}(x,y)= \Li_a(x) \Li_b(y)\,.  
\end{equation}
to give 
\begin{equation}
  L_{a,b;c}(x,y)={ 1 \over (c-1)! } \int_0^\infty d \alpha \,  
  \alpha^{c-1} \, \Li_a(x  e^{-\alpha})\Li_b(y e^{-\alpha})   \,.
\end{equation}
This  integral representation can then
be efficiently integrated with {\tt HyperInt}~\cite{Panzer:2014caa}.
The expression can be written in different ways as a  consequence of the
shuffle algebra. For example
\begin{equation}
  L_{1,1;1}(x,y)= \Li_{1,2}({y\over x},x)+\Li_{1,2}({x\over y},y)  \,,
\end{equation}
which is easily checked with {\tt HyperInt} to be  equal to 
\begin{equation}
L_{1,1;1}(x,y)= \Li_3(x)+ (\Li_2(x) - \Li_2(x\over y)) \Li_1(y) +
\Li_{1,2}(y,{x\over y})\,.
\end{equation}
Equating these two expressions evaluated at $x=y=1$  implies Euler's
famous relation $\zeta(1,2)=\zeta(3)$.

%%%%%%%%%%%%%%%%%%%%%%%%%%%%%%%%%%%%%%%%%%%%%%%%%%%%%%%%%%%%%%

\end{document}